\documentclass[submission, Phys]{SciPost}

\usepackage[export]{adjustbox}

\hypersetup{
    colorlinks,
    linkcolor={red!50!black},
    citecolor={blue!50!black},
    urlcolor={blue!80!black}
}

\usepackage[bitstream-charter]{mathdesign}
\DeclareSymbolFont{usualmathcal}{OMS}{cmsy}{m}{n}
\DeclareSymbolFontAlphabet{\mathcal}{usualmathcal}

\begin{document}

\begin{center}{\Large \textbf{
Critical behavior of a phase transition in the dynamics of interacting
populations\\
}}\end{center}

% TODO: write the author list here. Use first name (+ other initials) + surname format.
% Separate subsequent authors by a comma, omit comma and use "and" for the last author.
% Mark the corresponding author with a superscript star.
\begin{center}
Thibaut Arnoulx de Pirey\textsuperscript{1} and
Guy Bunin\textsuperscript{2}
\end{center}

\begin{center}
{\bf 1} Paris, France
\\
{\bf 2} Department of Physics, Technion-Israel Institute of Technology, Haifa 32000, Israel
\\
${}^\star$ {\small \sf thibautdepirey@gmail.com}
\end{center}

\begin{center}
\today
\end{center}

% For convenience during refereeing (optional),
% you can turn on line numbers by uncommenting the next line:
%\linenumbers

\section*{Abstract}
{\bf
% TODO: write your abstract here.
Many-variable differential equations with random coefficients provide
powerful models for the dynamics of many interacting species in ecology.
These models are known to exhibit a dynamical phase transition from
a phase where population sizes reach a fixed point, to a phase where
they fluctuate indefinitely. Here we provide a theory for the critical behavior close to the phase
transition. We show that timescales diverge at the transition and that temporal fluctuations
grow continuously upon crossing it. We further show the existence of three different universality classes, with different sets of critical exponents,
depending on the migration
rate which couples the system to its surroundings.
}

\vspace{10pt}
\noindent\rule{\textwidth}{1pt}
\tableofcontents\thispagestyle{fancy}
\noindent\rule{\textwidth}{1pt}
\vspace{10pt}

\section{Introduction}\label{sec:intro}

Many high-dimensional dynamical systems, of interest in neuroscience
\cite{sompolinsky1988chaos}, ecology \cite{opper1992phase,bunin2017ecological,galla2006random},
game theory \cite{galla2013complex} and economics \cite{dessertaine2022out},
exhibit two dynamical phases, one where the dynamics reach a fixed
point, and another where the variables fluctuate indefinitely. A sharp
phase transition between these phases is observed when varying system
parameters. In many cases, this transition has been linked to a loss
of fixed point stability and the emergence of unstable fixed points
whose number is exponentially large in the dimension of the system
\cite{BenArousFyodorov,ros2023generalized,PhysRevLett.110.118101}.

The first instance where such a transition was predicted are models
of high-dimensional random neural networks \cite{sompolinsky1988chaos},
which are by now well-understood \cite{crisanti2018path}. Here,
as for the p-spin dynamics \cite{crisanti1993spherical}, the dynamics
of a single degree of freedom become Gaussian in the high-dimensional
limit and closed equations for two-time correlation functions can
then be derived \cite{sompolinsky1981dynamic,sompolinsky1988chaos}.
In the fluctuating phase, the dynamics reach a chaotic time-translation
invariant state \cite{sompolinsky1988chaos}. The transition is of
second order, with a continuous growth of the amplitude of time fluctuations
when going away from the critical point. The critical behavior has
been investigated and the critical exponents governing the growth
of fluctuations and timescales have been obtained for various types
of nonlinearities \cite{kadmon2015transition}.

\begin{figure*}
\begin{centering}
\includegraphics[width=0.7\textwidth]{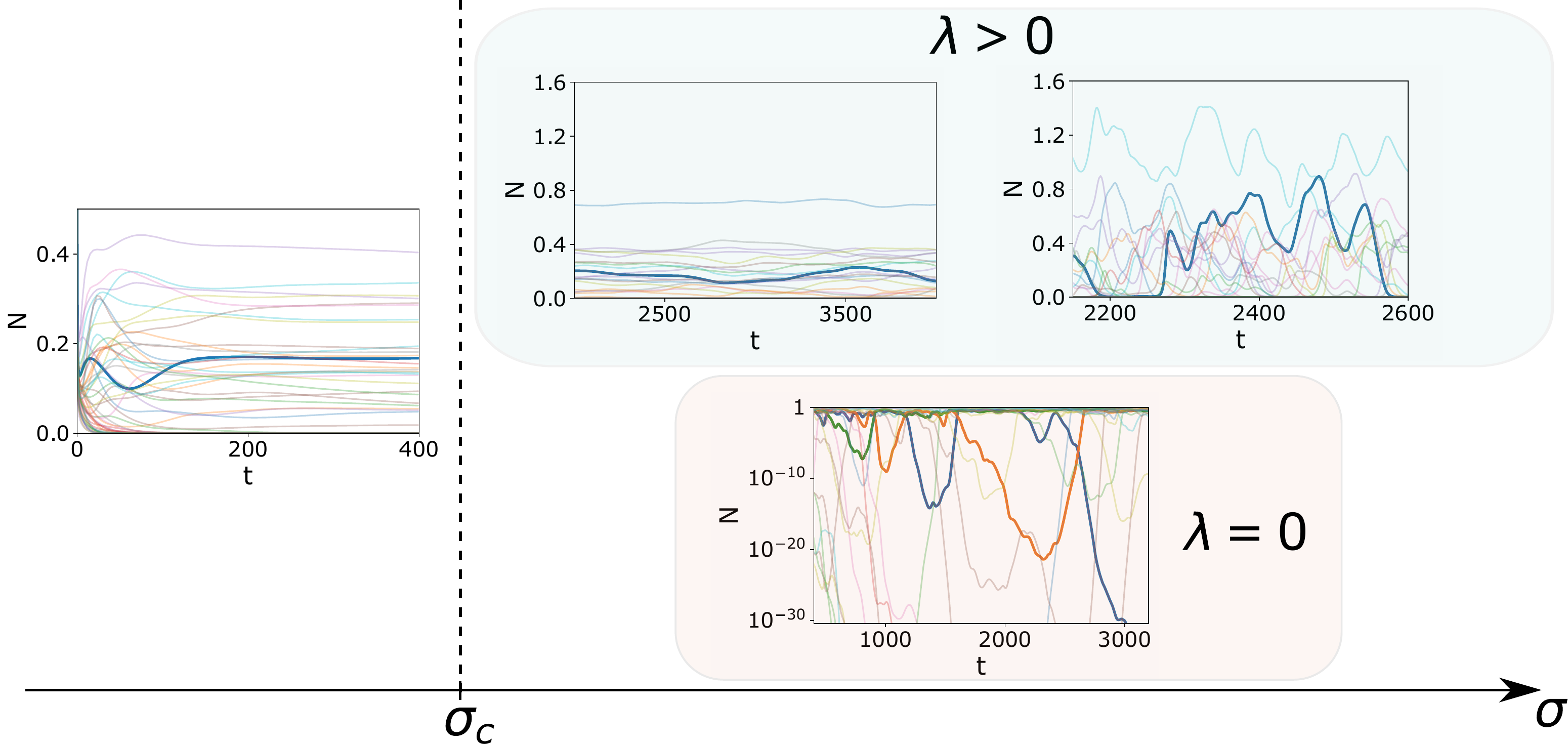}
\par\end{centering}
\caption{\label{fig:illustration}\textbf{Phases of the Lotka-Volterra dynamics.
}Example degrees of freedom from a simulation of the dynamics Eq.
(\ref{eq:many_bodyLV}) with $\mu=10$ and $S=4000$ for increasing
values of $\sigma$.\textbf{ }When\textbf{ $\sigma<\sigma_{c}$},
the dynamics reach a fixed point. Here $\sigma-\sigma_{c}=-0.3$.
When $\lambda>0$ and $\sigma>\sigma_{c}$ the dynamics reach a time-translation
invariant state. The amplitude of the time-fluctuations grows with
$\sigma$, and their correlation time decreases with $\sigma$. Here
$\lambda=10^{-8}$ and $\sigma-\sigma_{c}=0.1$ (Upper left) and $\sigma-\sigma_{c}=0.4$
(Upper right). When $\lambda=0$ and $\sigma>\sigma_{c}$, the dynamics
is aging, with growing fluctuations of log-populations sizes. Here
$\sigma-\sigma_{c}=0.4$.}
\end{figure*}

Species-rich ecosystems can also be modeled by high-dimensional nonlinear
dynamics for the different species population sizes. Common ecological
models, such as Lotka-Volterra or resource-competition models \cite{hofbauer1998evolutionary},
with random interactions between species, are also known to exhibit
a phase transition from fixed point to persistent fluctuations \cite{opper1992phase,bunin2017ecological}.
The Lotka-Volterra dynamics are a standard model for a well-mixed
system (no spatial extension). The dynamics of the population sizes
$N_{i}$ of the species $i=1\dots S$, with $S\gg1$ the number of
species read \cite{may2007theoretical,takeuchi1996global,barbier_Generic_2018}

\begin{equation}
\dot{N}_{i}=N_{i}\left(1-N_{i}-\sum_{j}\alpha_{ij}N_{j}\right)+\lambda\,.\label{eq:many_bodyLV}
\end{equation}
The matrix $\alpha_{ij}$ quantifies the interactions between species,
and $\lambda$ accounts for migration of individuals into the community
from its surroundings. We consider randomly sampled interaction matrices
with independent and identically-distributed entries, such that $\text{mean}(\alpha_{ij})=\mu/S$
and $\text{std}(\alpha_{ij})=\sigma/\sqrt{S}$. As the parameter $\sigma$
is increased and crosses a critical value $\sigma_{c}$, the system
exhibits a transition between a fixed point phase and a fluctuating
one \cite{bunin2017ecological,galla2018dynamically}, see Fig. \ref{fig:illustration}.
The biological relevance of such theoretical descriptions was demonstrated
experimentally in \cite{hu2022emergent}.

Yet, due to the multiplicative nature of population growth ($\dot{N}_{i}$
grows with $N_{i}$), ecological models can exhibit unique properties,
and since the dynamics of the population sizes $N_{i}(t)$ is non-Gaussian,
analytical predictions remain challenging. In sharp contrast to the
time-translational invariant state reached by other dynamical systems,
the dynamics Eq. (\ref{eq:many_bodyLV}) in the absence of migration
(i.e. when $\lambda=0$) exhibit \emph{aging}. Namely, the correlation
time increases with the elapsed time. This was shown numerically in
\cite{roy2019numerical}, and described analytically in \cite{de2024many},
where it was proven that the correlation time increases linearly with
the elapsed time. This aging behavior is characterized by variables
performing ever larger excursions to values near $N_{i}=0$, see Fig.
\ref{fig:illustration}, so that the system spends long times in the
vicinity of unstable fixed points. This aging is very different from
glassy dynamics on a rough landscape \cite{de2023aging}. In the
high-dimensional limit $S\to\infty$ and at long times, population
sizes follow non-Markovian jump-diffusion processes \cite{de2024many}.
At positive migration rate $\lambda>0$, the dynamics do reach a time-translation
invariant state \cite{roy2019numerical}, characterized by a correlation
time that grows as $|\ln\lambda|$ when $\lambda$ is small \cite{de2024many}.
The origin of the long timescale can be traced back to population
sizes growing and declining exponentially between $O(\lambda)$ and
$O(1)$ values. Figure \ref{fig:illustration} recapitulates these
different phases of the Lotka-Voltera dynamics.

Understanding the behavior of these dynamics in the critical regime
close to the transition, and how they are affected by the phase space
boundaries at $N_{i}=0$, has so far remained an open problem. Earlier
numerical work showed that, as the transition is approached, the amplitude
of the fluctuations gets smaller, and timescales grow, but the precise
scaling behavior has not been elucidated \cite{roy2019numerical}.

\begin{figure}
\begin{centering}
\includegraphics[width=1\columnwidth]{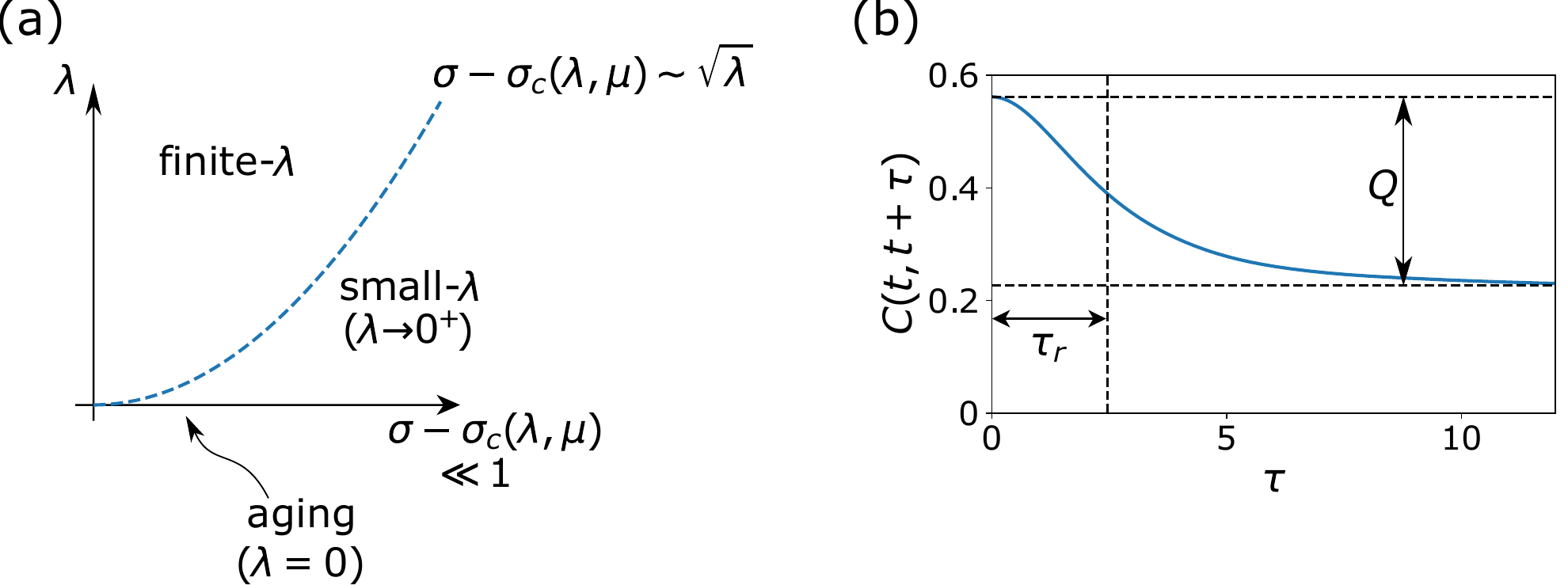}
\par\end{centering}
\caption{\label{fig:regimes}\textbf{Critical regimes of the Lotka-Volterra
dynamics. (a)} Three different scaling regimes are identified when
$\sigma-\sigma_{c}$ is small: when $\lambda=0$; when $\lambda>0$
and fixed; and in the limit $\lambda\to0^{+}$, taken before the limit
$\sigma\to\sigma_{c}^{+}$. The crossover between the second and third
regimes takes place at $\sigma-\sigma_{c}\sim\sqrt{\lambda}.$ \textbf{(b)}
Two-time autocorrelation function $C(t,t')\equiv\sum_{i}N_{i}(t)N_{i}(t')/S$
in the fluctuating phase, with definitions of the order parameters
$Q$ (the amplitude of the fluctuations) and $\tau_{r}$ (the relaxation
time of the fluctuations). As the critical point is approached from
above, $\sigma\to\sigma_{c}^{+}$, $Q$ decreases and $\tau_{r}$
increases. Here $C(t,t')$ is shown for $\lambda=0.1$, $\mu=10$
and $\sigma-\sigma_{c}=0.1$. }
\end{figure}

In this work, we provide a comprehensive analytical description of
the critical regime of the Lotka-Volterra dynamics Eq. (\ref{eq:many_bodyLV})
when $\sigma$ goes to $\sigma_{c}$ from above, meaning from inside
the fluctuating phase. We show that there exist three universality
classes depending on the relative size of the migration rate $\lambda$,
and the distance to the transition $\sigma-\sigma_{c}$, see Fig.
\ref{fig:regimes}(a). One scaling regime is obtained when $\lambda=0$,
where the transition point $\sigma=\sigma_{c}$ separates a fixed
point phase to an aging phase. Another scaling regime is obtained
for $\lambda>0$ fixed, where the transition is from the fixed point
phase to a time-translation invariant chaotic phase. This regime is
characterized by the same critical exponents as random neural network
models with a strongly non-linear transfer function \cite{kadmon2015transition}.
A third scaling regime is obtained when both $\lambda\to0^{+}$ and
$\sigma-\sigma_{c}\to0^{+}$, while keeping $\sigma-\sigma_{c}\gg\sqrt{\lambda}$.
For each regime, we describe the dynamics near criticality in terms
of a scaling theory for the growth of fluctuations and timescales,
and derive the corresponding critical exponents. We find the position
of the critical point $\sigma_{c}(\lambda,\mu)$ and show that the
chaotic phase does not exist for large values of $\lambda$.

The paper is organized as follows. We present a summary of our main
results in Sec. \ref{sec:Summary-of-the}. The rest of the paper is
dedicated to the derivation of these results. In Sec. \ref{sec:DMFT},
we recall the Dynamical Mean Field Theory (DMFT) equations associated
to the Lotka-Volterra dynamics, which form the basis of our analysis.
In Sec. \ref{sec:RescaledDyn}, we discuss the two cases $\lambda=0$
and $\lambda\to0^{+}$. Then, in Sec. \ref{sec:finite_l}, we consider
the case where the migration rate $\lambda>0$ is fixed and positive.

\section{Summary of the main results}\label{sec:Summary-of-the}

In this section, we summarize the main results for each of the universality
classes for the main two quantities of interest: The amplitude of
the fluctuations $Q$ and the associated relaxation timescale $\tau_{r}$,
see Fig. \ref{fig:regimes} (b).

\subsection{Growth of fluctuations}

\begin{figure*}
\begin{centering}
\includegraphics[width=1\textwidth]{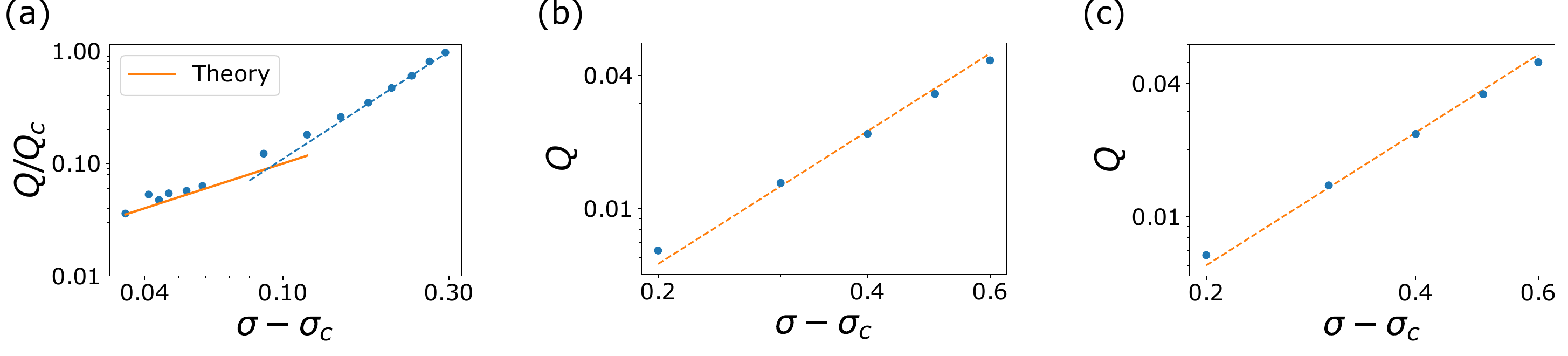}
\par\end{centering}
\caption{\label{fig:growth_fluctuations}\textbf{Growth of fluctuations in
the vicinity of the critical point.} \textbf{(a)} $\lambda>0$ criticality.
For small $\sigma-\sigma_{c}(\lambda,\mu)$ the amplitude of the fluctuations
is predicted to grow to leading order as $Q/Q_{c}=(\sigma-\sigma_{c})$
with $Q_{c}$ a known function of $\lambda$ and $\mu$. The dots
are obtained from numerically solving the DMFT equations of Sec. \ref{sec:DMFT}.
The solid line is the analytical prediction $Q/Q_{c}=(\sigma-\sigma_{c})$.
This is a parameter-free agreement, as $Q_{c}$ is here known analytically.
At larger values of $\sigma-\sigma_{c}(\lambda,\mu)$ the scaling
becomes quadratic, as shown by a fit of the form $Q/Q_{c}=a\,(\sigma-\sigma_{c})^{2}$
(dashed line). Here $\lambda=0.1$ and $\mu=10$. \textbf{(b) $\lambda=0$
}criticality. For small $\sigma-\sigma_{c}(\lambda=0,\mu)$ the amplitude
of the fluctuations is predicted to grow as $Q\sim(\sigma-\sigma_{c})^{2}$.\textbf{
}The dots are obtained from numerically solving the rescaled DMFT
equations of Sec. \ref{subsec:Mapping-nomig} and the dashed line
is a fit of parameter $Q_{c}$ to the form $Q=Q_{c}\,(\sigma-\sigma_{c})^{2}$.
Here $\mu=10$. \textbf{(c) $\lambda\to0^{+}$ }criticality. For small
$\sigma-\sigma_{c}(\lambda=0,\mu)$ the amplitude of the fluctuations
is predicted to grow as $Q\sim(\sigma-\sigma_{c})^{2}$.\textbf{ }The
dots are obtained from numerically solving the rescaled DMFT equations
of Sec. \ref{subsec:Mapping-l0+} and the dashed line is a fit of
parameter $Q_{c}$ to the functional form $Q=Q_{c}\,(\sigma-\sigma_{c})^{2}$.
Here $\mu=10$.}
\end{figure*}

For fixed parameters $\lambda\geq0$ and $\mu$, the system is in
a fluctuating phase when $\sigma>\sigma_{c}(\lambda,\mu)$, with $\sigma_{c}(\lambda,\mu)\to\sqrt{2}$
as $\lambda\to0$. We study the autocorrelation function of the population
sizes $C(t,t')\equiv\sum_{i}N_{i}(t)N_{i}(t')/S$, in the large $S$
limit and for long times $t,t'$, from which we extract the amplitude
and relaxation time of the fluctuations, see Fig. \ref{fig:regimes}
(b). The autocorrelation function fully characterizes the long-time
dynamics of the population sizes, as explained in Sec. \ref{sec:DMFT}.
The amplitude $Q$ of the fluctuations is defined from $C(t,t')$
through
\begin{equation}
Q=\lim_{t\to\infty}\left(C(t,t)-\lim_{\tau\to\infty}C(t,t+\tau)\right)\,,\label{eq:defQ}
\end{equation}
which grows continuously from 0 at the transition, with an exponent
$\beta$ defined by
\begin{equation}
Q\sim Q_{c}(\lambda,\mu)|\sigma-\sigma_{c}(\lambda,\mu)|^{\beta}\,.\label{eq:defQc}
\end{equation}
We obtain the following results in the three regimes of interest.
First, when the limit $\lambda\to0^{+}$ is taken before the limit
$\sigma\to\sigma_{c}(0,\mu)=\sqrt{2}$ and when $\lambda=0$, we obtain
$\beta=2$, see Sec. \ref{sec:Growth-of-fluctuations-1} and Sec.
\ref{sec:Growth-of-fluctuations-2}. However when $\lambda>0$ is
fixed, we obtain $\beta=1$, see Sec. \ref{sec:finite_l}, Eq. (\ref{eq:condition}).
We also find an analytical expression for the coefficient $Q_{c}(\lambda,\mu)$
in that case, see Sec. \ref{sec:amplitude_finite_l}. We confirm these
predictions in numerical solutions of the DMFT equations established
in Sec. \ref{sec:DMFT}, see Fig. \ref{fig:growth_fluctuations}.

\subsection{Critical slowing down}

At long times, the autocorrelation function $C(t,t')$ can be written
as a sum of a steady part and a transient part $\delta C(t,t')$,
\begin{equation}
C(t,t')=w^{2}+Q\,\delta C(t,t')\label{eq:defdeltaC}
\end{equation}
with $\delta C(t,t)=1$ and $\lim_{\tau\to\infty}\delta C(t,t+\tau)=0$.
If the limit $\lambda\to0^{+}$ is taken at fixed $\sigma>\sqrt{2}$,
the dynamics reach a time-translation invariant state with a long
correlation time proportional to $|\ln\lambda|$, meaning that the
dynamics become regular as $\lambda\to0^{+}$ when described over
timescales of order $O(|\ln\lambda|)$ \cite{de2024many}. When $|\sigma-\sqrt{2}|\ll1$,
with still $\sigma-\sqrt{2}\gg\sqrt{\lambda}$, the dynamics are described
by a critical regime described by the scaling form
\begin{equation}
\delta C(t,t')\sim\delta\hat{C}_{+}\left(\frac{|t-t'|/|\ln\lambda|}{|\sigma-\sqrt{2}|^{-\zeta}}\right)\,,\label{eq:correl_scaling_l0}
\end{equation}
with $\zeta=1$ and $\delta\hat{C}_{+}$ a scaling function, see Sec.
\ref{sec:Growth-of-fluctuations-1}. 

When $\lambda=0$ the dynamics in the fluctuating phase exhibits an
aging behavior with a correlation time that grows linearly with the
elapsed time \cite{de2024many}. In other words, the dynamics is
time-translation invariant in log-time ($\ln t$). We show that in
log-time there is no critical slowing down, meaning that close to
the transition
\begin{equation}
\delta C(t,t')\sim\delta\hat{C}_{0}\left(\frac{|\ln t-\ln t'|}{|\sigma-\sqrt{2}|^{-\zeta}}\right)\,,\label{eq:correl_nomig}
\end{equation}
with $\zeta=0$ and $\delta\hat{C}_{0}$ another scaling function,
see Sec. \ref{sec:Growth-of-fluctuations-2}.

\begin{figure}
\begin{centering}
\includegraphics[width=0.8\columnwidth]{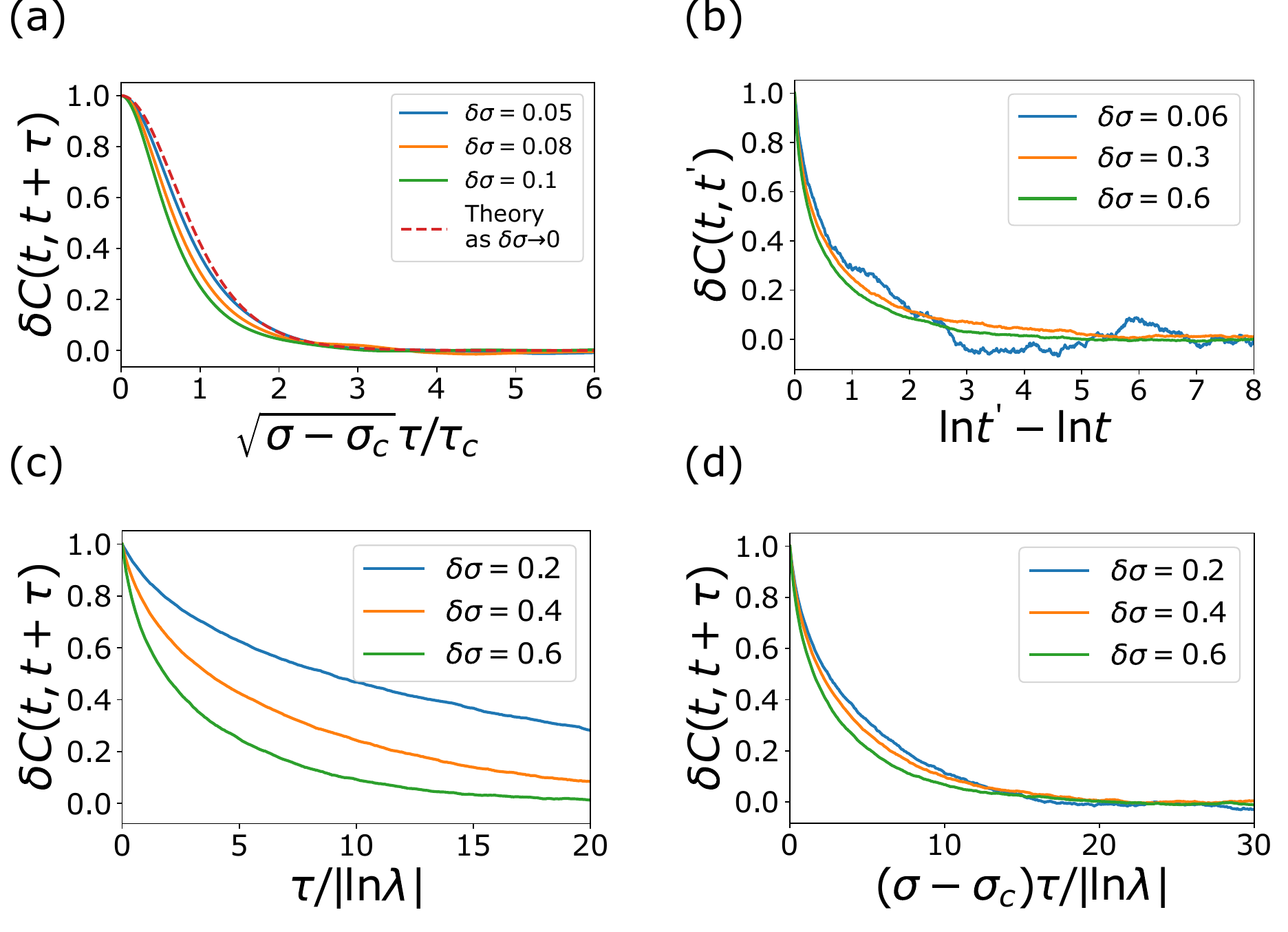}
\par\end{centering}
\caption{\label{fig:correlation_function}\textbf{Growth of timescales in the
vicinity of the critical point.} \textbf{(a)} $\lambda>0$ criticality.
For small $\delta\sigma=\sigma-\sigma_{c}(\lambda,\mu)$, the rescaled
correlation function $\delta C(t,t+\tau)$ converges to the scaling
form given in Eq. (\ref{eq:correl_lpos}). The continuous lines are
obtained by numerically solving the DMFT equation of Sec. \ref{sec:DMFT}.
This is a parameter-free agreement: the dashed red line is the analytical
prediction of Eq. (\ref{eq:correl_lpos}), and $\tau_{c}(\lambda,\mu)$
is known. Here $\lambda=0.1$ and $\mu=10$. \textbf{(b)} $\lambda=0$
criticality. For small $\delta\sigma=\sigma-\sqrt{2}$, the rescaled
correlation function converges to the scaling form of Eq. (\ref{eq:correl_nomig}).
Note the collapse of the correlation functions for different values
of $\delta\sigma$ without rescaling of the log-time axis. The continuous
lines are obtained by numerically solving the DMFT equation of Sec.
\ref{subsec:Mapping-nomig}. Here $\mu=10$.\textbf{ (c, d)} $\lambda\to0^{+}$
criticality. \textbf{(c)} The correlation time of the fluctuations
diverge as $\sigma\to\sqrt{2}$ from above. \textbf{(d)} For small
$\delta\sigma=\sigma-\sqrt{2}$, the timescale in rescaled time grows
as $(\sigma-\sqrt{2})^{-1}$. The continuous lines are obtained by
numerically solving the DMFT equation of Sec. \ref{subsec:Mapping-l0+}.
Here $\mu=10$.}
\end{figure}

Lastly, when $\sigma>\sigma_{c}(\lambda,\mu)$ and $\lambda>0$ fixed,
the dynamics reach a time-translation invariant state. When approaching
the critical point $\sigma\to\sigma_{c}(\lambda,\mu)$ from above
while keeping $\lambda$ fixed, the relaxation time of the fluctuations
grows with an exponent $\zeta$ inferred from the scaling form 
\begin{equation}
\delta C(t,t')\sim1-\text{Tanh}^{2}\left(\frac{|t-t'|}{\tau_{c}(\lambda,\mu)|\sigma-\sigma_{c}(\lambda,\mu)|^{-\zeta}}\right)\,.\label{eq:correl_lpos}
\end{equation}
We show that $\zeta=1/2$, see the discussion after Eq. (\ref{eq:condition})
in Sec. \ref{sec:finite_l}. The scaling function, as well as the
value of the parameter $\tau_{c}(\lambda,\mu)$, are obtained in Eqs.
(\ref{eq:result_C}) and (\ref{eq:taucres}) respectively. We confirm
these predictions in numerical solutions of the Dynamical Mean Field
Theory equations established in Sec. \ref{sec:DMFT}, see Fig. \ref{fig:correlation_function}.

Note that the correlation functions $\delta\hat{C}_{+}(x)$ and $\delta\hat{C}_{0}(x)$
are non-differentiable at $x=0$, namely the slope $d\left(\delta\hat{C}_{+}\right)/dx$
has different values at $x\to0^{+}$ and $x\to0^{-}$. This means
that trajectories in rescaled time have a Brownian motion component
\cite{de2024many}. In contrast, at $\lambda>0$, $\delta C(t,t')$
is differentiable at $t=t'$, meaning that trajectories do not have
a Brownian motion component.

\subsection{Crossover between the finite $\lambda>0$ and the $\lambda\to0^{+}$
critical behaviors\label{subsec:crossover}}

For $\lambda\ll1$, the shift in the critical point is proportional
to leading order to $\sqrt{\lambda}$, $\sigma_{c}(\lambda,\mu)-\sqrt{2}\sim\sqrt{\lambda}$.
The amplitude $Q_{c}(\lambda,\mu)$, see Eq. (\ref{eq:defQc}), setting
the amplitude of the critical fluctuations beyond the exponent $\beta$,
also goes to 0 as
\[
Q_{c}(\lambda,\mu)\sim\sqrt{\lambda}\,,
\]
see Eq. (\ref{eq:qc_smalll}) in Sec. \ref{sec:amplitude_small_l}.
At the same time, the timescale $\tau_{c}(\lambda,\mu)$ in Eq. (\ref{eq:correl_lpos})
increases with
\[
\tau_{c}(\lambda,\mu)\sim\lambda^{-1/4}
\]
see Eq. (\ref{eq:correl_time_small}) in Sec. \ref{sec:amplitude_small_l}.
Combining this with Eqs. (\ref{eq:correl_scaling_l0}) and recalling
that the exponent in Eq. (\ref{eq:defQc}) is $\beta=2$ when $\lambda\to0^{+}$,
one finds a crossover at $\sigma-\sqrt{2}\sim\sqrt{\lambda}$ when
both $\lambda\ll1$ and $\sigma-\sqrt{2}\ll1$. The finite $\lambda>0$
critical regime dominates for $\sigma-\sqrt{2}\ll\sqrt{\lambda}$
while the $\lambda\to0^{+}$ critical regime dominates when $\sigma-\sqrt{2}\gg\sqrt{\lambda}$.
The crossover is illustrated in Fig. \ref{fig:growth_fluctuations}
(a).

\section{Dynamical mean-field theory\label{sec:DMFT}}

Our theory is built on dynamical mean-field theory (DMFT). Originally
developed in the context of spin-glasses \cite{de1978dynamics},
and later adapted to ecological dynamics in \cite{opper1992phase}
and to the present equations in \cite{galla2018dynamically,roy2019numerical},
it shows that in the limit $S\to\infty$ and for $N_{i}$ sampled
independently at the initial time, the dynamics of the population
sizes are described by independent realizations of the stochastic
differential equation

\begin{equation}
\dot{N}(t)=N(t)\left[1-N(t)-\mu m(t)+\sigma\xi(t)\right]+\lambda\,,\label{eq:DMFT}
\end{equation}
where the index $i$ has been dropped, and where $\xi(t)$ is a zero-mean
Gaussian process and $m(t)$ a deterministic function of time. This
can be shown to come from the fact that the term $\xi_{i}(t)\equiv\sum_{j}\alpha_{ij}N_{j}(t)$
in Eq. (\ref{eq:many_bodyLV}) is the sum of many weakly-correlated
contributions. However, $m(t)$ and the correlations of $\xi(t)$
are not provided in advance. Instead, they are derived through self-consistency
conditions. This is a dynamical equivalent of the self-consistency
condition on the magnetization derived in the mean-field Ising model,
for instance. Recalling that $\langle\alpha_{ij}\rangle=\mu/S$ and
$\langle\alpha_{ij}\alpha_{kl}\rangle-\langle\alpha_{ij}\rangle\langle\alpha_{kl}\rangle=\sigma^{2}\delta_{ik}\delta_{kl}/S$,
the self-consistency conditions read 
\begin{equation}
m(t)=\left\langle N(t)\right\rangle \,,\label{eq:DMFT1-1}
\end{equation}

\noindent and 
\begin{equation}
\left\langle \xi(t)\xi(t')\right\rangle =\left\langle N(t)N(t')\right\rangle \,,\label{eq:DMFT2}
\end{equation}

\noindent where the average $\left\langle \dots\right\rangle $ stands
for an average over the stochastic process in (\ref{eq:DMFT}). Because
$\xi(t)$ is a Gaussian process, obtaining the first two moments of
the population size $\left\langle N(t)\right\rangle $ and $\left\langle N(t)N(t')\right\rangle $
allows to completely characterize the dynamics of $N(t)$ by using
Eq. (\ref{eq:DMFT}).

\noindent We denote by $m_{\infty}$ the steady-state value of $m(t)$.
At long-times, we further decompose the noise $\xi(t)$ into a frozen
Gaussian random variable $\xi_{\infty}$ and a Gaussian process $\delta\xi(t)$
that completely decorrelates over time, meaning $\xi(t)\equiv\xi_{\infty}+\epsilon\delta\xi(t)$
with $\lim_{\tau\to\infty}\left\langle \delta\xi(t)\delta\xi(t+\tau)\right\rangle =0$.
The amplitude $\epsilon$ of the fluctuating part is defined so that
$\left\langle \delta\xi(t)^{2}\right\rangle =1$. Using the self-consistency
condition Eq. (\ref{eq:DMFT2}), we get from the decomposition in
Eq. (\ref{eq:defdeltaC}) that $\left\langle \xi_{\infty}^{2}\right\rangle =w^{2}$
and $\left\langle \delta\xi(t)\delta\xi(t+\tau)\right\rangle =\delta C(t,t+\tau)$.
We also obtain $Q=\epsilon^{2}$. To make the notations more compact,
we introduce $g\equiv(1-\mu m_{\infty})/\sigma+\xi_{\infty}$ and
$\tilde{m}\equiv(1-\mu m_{\infty})/\sigma$. The dynamics in Eq. (\ref{eq:DMFT})
then becomes at long-times
\begin{equation}
\dot{N}(t)=N(t)\left[\sigma(g+\epsilon\delta\xi(t))-N(t)\right]+\lambda\,.\label{eq:DMFT_redef}
\end{equation}

\noindent The random variable $g$ is Gaussian distributed with yet
unknown mean and variance. Its distribution is denoted $P(g)$ and
is given by 
\begin{equation}
P(g)=\frac{1}{\sqrt{2\pi}w}\exp\left(-\frac{\left(g-\tilde{m}\right)^{2}}{2w^{2}}\right)\,.\label{eq:distrib_g}
\end{equation}

\noindent We recall that the average $\left\langle \dots\right\rangle $
runs over realizations of the frozen variable $g$ and the Gaussian
process $\delta\xi(t)$. Below it will be convenient to perform this
average in two parts. We thus introduce the notation $\left\langle \dots\right\rangle _{g}$
to denote an average over the Gaussian process $\delta\xi(t)$ at
fixed $g$. Equations (\ref{eq:DMFT1-1},\ref{eq:DMFT2},\ref{eq:DMFT_redef},\ref{eq:distrib_g})
form the basis of the subsequent analysis.

\section{Critical regimes when $\lambda\to0^{+}$ and $\lambda=0$\label{sec:RescaledDyn}}

We now determine the properties of the dynamical phase transition
in the two cases where the limit $\lambda\to0^{+}$ is taken before
the limit $\sigma\to\sqrt{2}$ and where $\lambda=0$. As discussed
in Sec. \ref{subsec:crossover}, this first regime is relevant when
$\lambda\ll1$, $\sigma-\sqrt{2}\ll1$ and $\lambda\ll\sigma-\sqrt{2}$.
In those two cases, for any $\sigma>\sqrt{2}$ finite, it was shown
in \cite{de2024many}, that the dynamics evolve over infinitely long
timescales. 

When $\lambda\to0^{+}$, the dynamics is time-translation invariant
at long times with a long correlation time proportional to $|\ln\lambda|$,
with

\[
\lim_{\lambda\to0^{+}}\delta C_{\lambda}(t,t+\vert\ln\lambda\vert s)\to\delta\hat{C}_{+}(s)\,,
\]
with $\delta\hat{C}_{+}(s)$ a regular function as $s\to0^{+}$. 

In the absence of migration ($\lambda=0$),
\[
\lim_{t\to\infty}\delta C_{\lambda=0}(t,t\text{e}^{s})=\delta\hat{C}_{0}(s)\,,
\]
with $\delta\hat{C}_{0}(s)$ another regular function as $s\to0^{+}$.
This is typical of an aging dynamics, with a growth of timescale proportional
to the age of the system. Leveraging on these two identities, effective
equations for the long-time dynamics of the population size $N(t)$
were obtained in \cite{de2024many}. These long-time effective dynamics
are reviewed in Sec. \ref{subsec:Mapping-l0+} and Sec. \ref{subsec:Mapping-nomig},
for dynamics with $\lambda\to0^{+}$ and $\lambda=0$ respectively.
We then use these results to derive the corresponding scaling behaviors,
in Sec. \ref{sec:Growth-of-fluctuations-1} and Sec. \ref{sec:Growth-of-fluctuations-2}.

\subsection{Steady-state dynamics when $\lambda\to0^{+}$\label{subsec:Mapping-l0+}}

\begin{figure}
\begin{centering}
\includegraphics[width=0.5\columnwidth]{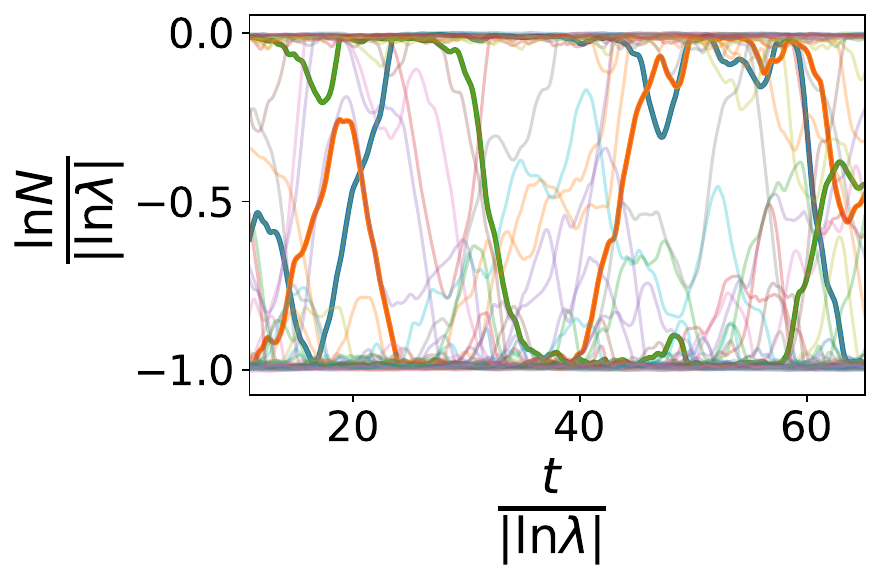}
\par\end{centering}
\caption{\label{fig:small_l_fig}\textbf{Slow dynamics in the fluctuating phase
for $\lambda\ll1$.} Example degrees of freedom from a simulation
of the dynamics Eq. (\ref{eq:many_bodyLV}) with $\lambda=10^{-40}$,
$\mu=10$, $\sigma-\sigma_{c}=0.4$ and $S=4000$.\textbf{ }The process
$z(s)$ with $s=t/|\ln\lambda|$ and $z=\ln N/|\ln\lambda|$ converges
when $\lambda\to0^{+}$ to a time-translation invariant process, confined
between $0$ and $-1$, with finite correlation time.}
\end{figure}

When $\lambda\to0^{+}$ and $\sigma-\sqrt{2}>0$ fixed, the effective
steady-state dynamics is described by a rescaling of the original
DMFT equations Eq. (\ref{eq:DMFT_redef}), as illustrated in Fig.
\ref{fig:small_l_fig}. We introduce $z=\ln N/|\ln\lambda|$ and $s=t/|\ln\lambda|$.
When $\lambda\to0^{+}$, the process $z(s)$ follows a well-defined
stochastic differential equation \cite{de2024many},
\begin{equation}
z'(s)=\sigma g+\sigma\epsilon\delta\hat{\xi}(s)-W(z)+W(-z-1)\label{eq:EOM_z_0+}
\end{equation}
with $\delta\hat{\xi}(s)\equiv\delta\xi(t)$ a zero-mean Gaussian
noise with correlations $\delta\hat{C}_{+}(s)$. The function $W(z)$
acts as a hard wall with $W(z>0)=+\infty$ and $W(z<0)=0$, so that
the dynamics are confined between $-1\leq z\leq0$. Equation (\ref{eq:EOM_z_0+})
is supplemented by an expression for $N(s)$ valid at long times,
\begin{equation}
N(s)=\sigma\,\Theta(z(s))\left(g+\epsilon\delta\hat{\xi}(s)\right)\,,\label{eq:EOM_N_l0+}
\end{equation}
where the Heaviside function $\Theta$ is used with the convention
$\Theta(0)=1$. Therefore, the system of self-consistency equations
in the $\lambda\to0^{+}$ regime reads
\begin{equation}
\frac{1-\sigma\tilde{m}}{\mu}=\sigma\left\langle \Theta(z(s))\left(g+\epsilon\delta\hat{\xi}(s)\right)\right\rangle \,,\label{eq:self-cons_longtime1}
\end{equation}
and
\begin{equation}
w^{2}+\epsilon^{2}\delta\hat{C}_{+}(s)=\sigma^{2}\left\langle \Theta(z(s))\Theta(z(0))\left(g+\epsilon\delta\hat{\xi}(s)\right)\left(g+\epsilon\delta\hat{\xi}(0)\right)\right\rangle \,,\label{eq:self-cons_longtime2}
\end{equation}
where all averages are performed in steady-state.

\subsection{Growth of fluctuations and timescales when $\lambda\to0^{+}$\label{sec:Growth-of-fluctuations-1}}

Equation (\ref{eq:self-cons_longtime2}), together with the dynamics
in Eq. (\ref{eq:EOM_z_0+}), gives us the self-consistency equation
satisfied by the correlation function $\delta\hat{C}_{+}(s)$. In
the right-hand side of Eq. (\ref{eq:self-cons_longtime2}), we split
the average over $g$ between (i) $g>0$, in which case $\Theta(z)=1$
with high probability for small $\epsilon$, and (ii) $g<0$ in which
case $\Theta(z)=0$ with high probability for small $\epsilon$. We
thus rewrite Eq. (\ref{eq:self-cons_longtime2}) as
\begin{align}
w^{2}+\epsilon^{2}\delta\hat{C}_{+}(s) & =\sigma^{2}\int_{0}^{+\infty}{\rm d}g\,P(g)\left(g^{2}+\epsilon^{2}\delta C_{+}(s)\right) \nonumber \\ & +\sigma^{2}\int_{-\infty}^{+\infty}{\rm d}g\,P(g)\left\langle \left[\Theta(z(s))\Theta(z(0))-\Theta(g)\right]\left(g+\epsilon\,\delta\xi(0)\right)\left(g+\epsilon\,\delta\xi(s)\right)\right\rangle _{g}\,.
\end{align}
No approximation is made in this rewriting. The first integral on
the right-hand side of the above equation can be computed using the
definition of $P(g)$ in Eq. (\ref{eq:distrib_g}), so the above equation
becomes
\begin{align*}
& \left(1-\frac{\sigma^{2}}{2}\right)\left(w^{2}+\epsilon^{2}\delta\hat{C}_{+}(s)\right)+\frac{\sigma^{2}}{2}\left(w^{2} +\epsilon^{2}\delta\hat{C}_{+}(s)\right){\rm Erf}\left(\frac{\tilde{m}}{\sqrt{2}w}\right) \\ & +\frac{\sigma^{2}}{2}\tilde{m}^{2}\left[1+{\rm Erf}\left(\frac{\tilde{m}}{\sqrt{2}w}\right)\right]+\frac{\tilde{m}w}{\sqrt{2\pi}}{\rm exp}\left(-\frac{\tilde{m}^{2}}{2w^{2}}\right)=\epsilon^{3}P(g=0)I(s)\,,
\end{align*}
where we introduced
\[
I(s)\equiv\frac{\sigma^{2}}{\epsilon^{3}P(g=0)}\int_{-\infty}^{+\infty}{\rm d}g\,P(g)\left\langle \left[\Theta(z(s))\Theta(z(0))-\Theta(g)\right]\left(g+\epsilon\,\delta\xi(0)\right)\left(g+\epsilon\,\delta\xi(s)\right)\right\rangle _{g}\,.
\]
We recall that $\delta\hat{C}_{+}(\infty)=0$, so that
\begin{equation}
\left(1-\frac{\sigma^{2}}{2}\right)w^{2}+\frac{\sigma^{2}}{2}w^{2}{\rm Erf}\left(\frac{\tilde{m}}{\sqrt{2}w}\right)+\frac{\sigma^{2}}{2}\tilde{m}^{2}\left[1+{\rm Erf}\left(\frac{\tilde{m}}{\sqrt{2}w}\right)\right]+\frac{\tilde{m}w}{\sqrt{2\pi}}{\rm exp}\left(-\frac{\tilde{m}^{2}}{2w^{2}}\right)=\epsilon^{3}P(g=0)I(\infty)\,.\label{eq:var_rescaled}
\end{equation}

\noindent Furthermore, $\delta\hat{C}_{+}(0)=1$ so that we get
\begin{equation}
\epsilon^{-1}\left(1-\frac{\sigma^{2}}{2}+\frac{\sigma^{2}}{2}{\rm Erf}\left(\frac{\tilde{m}}{\sqrt{2}w}\right)\right)=P(g=0)\left(I(0)-I(\infty)\right).\label{eq:amplitude}
\end{equation}

\noindent Hence we obtain an exact nonlinear equation satisfied by
the correlation function 
\begin{equation}
\delta\hat{C}_{+}(s)=\frac{I(s)-I(\infty)}{I(0)-I(\infty)}.\label{eq:self_consdC_l0+}
\end{equation}
We can now study this equation when $\epsilon\ll1$. In that case,
the contribution to $I(s)$ coming from $O\left(\epsilon^{0}\right)$
values of $g$ are exponentially small in $1/\epsilon$. The only
perturbative contributions to $I(s)$ arise from the values $g=O(\epsilon)$.
We therefore introduce $\tilde{g}\equiv g/\epsilon$. Accordingly,
we rescale time as $\tilde{s}=\epsilon s$ and denote $\tilde{z}(\tilde{s})\equiv z(\tilde{s}/\epsilon)$
which satisfies an equation like that for the original process $z(s)$,
Eq. (\ref{eq:EOM_z_0+}), but with an $O(1)$ fluctuating noise,
\begin{equation}
\frac{{\rm d}\tilde{z}}{{\rm d}\tilde{s}}=\sigma\left(\tilde{g}+\delta\xi(\tilde{s})\right)-W(\tilde{z})+W(-\tilde{z}-1)\,.\label{eq:process_ztile}
\end{equation}
In terms of this process, we get
\begin{equation}
I(\tilde{s})=\frac{\sigma^{2}}{P(g=0)}\int_{-\infty}^{+\infty}{\rm d}\tilde{g}\,P(\epsilon\tilde{g})\left\langle \left[\Theta(\tilde{z}(\tilde{s}))\Theta(\tilde{z}(0))-\Theta(\tilde{g})\right]\left(\tilde{g}+\delta\xi(0)\right)\left(\tilde{g}+\delta\xi(\tilde{s})\right)\right\rangle _{\tilde{g}}\,\label{eq:Is_rescaled}
\end{equation}
Hence $I=O\left(\epsilon^{0}\right)$, and to leading order
\begin{equation}
I(\tilde{s})=\sigma^{2}\int_{-\infty}^{+\infty}{\rm d}\tilde{g}\left\langle \left[\Theta(\tilde{z}(\tilde{s}))\Theta(\tilde{z}(0))-\Theta(\tilde{g})\right]\left(\tilde{g}+\delta\xi(0)\right)\left(\tilde{g}+\delta\xi(\tilde{s})\right)\right\rangle _{\tilde{g}}+O\left(\epsilon\right)\,,\label{eq:Is_expansion}
\end{equation}
as $P(\epsilon\tilde{g})$ can be replaced by $P(0)$. A similar reasoning
can be applied to the first moment equation, Eq. (\ref{eq:self-cons_longtime1}),
and yields,
\begin{equation}
\left(\frac{1}{\mu}-\frac{\sigma w}{\sqrt{2\pi}}\right)-\frac{\sigma\tilde{m}}{\mu}-\frac{\sigma w}{\sqrt{2\pi}}\left[{\rm exp}\left(-\frac{\tilde{m}^{2}}{2w^{2}}\right)-1\right]-\frac{\sigma\tilde{m}}{2}\left[1+{\rm Erf}\left(\frac{\tilde{m}}{\sqrt{2}w}\right)\right]=\epsilon^{2}P(g=0)J\,,\label{eq:mean_rescaled}
\end{equation}
with
\begin{align*}
J & =\frac{\sigma}{P(g=0)}\int_{-\infty}^{+\infty}{\rm d}\tilde{g}\,P(\epsilon\tilde{g})\left\langle \left[\Theta(\tilde{z}(\tilde{s}))-\Theta(\tilde{g})\right]\left(\tilde{g}+\delta\xi(\tilde{s})\right)\right\rangle _{\tilde{g}}\,,\\
 & =\sigma\int_{-\infty}^{+\infty}{\rm d}\tilde{g}\left\langle \left[\Theta(\tilde{z}(\tilde{s}))-\Theta(\tilde{g})\right]\left(\tilde{g}+\delta\xi(\tilde{s})\right)\right\rangle _{\tilde{g}}+O\left(\epsilon\right)\,.
\end{align*}

\noindent We can now obtain the scaling behavior close to the critical
point. Equation (\ref{eq:self_consdC_l0+}) provides a nonlinear equation
for the leading order correlation function $\delta\hat{C}_{+}(\tilde{s})$,
when $I(\tilde{s})$ is replaced by its leading order expression in
Eq. (\ref{eq:Is_expansion}). This equation cannot be solved explicitly
since the average entering Eq. (\ref{eq:Is_expansion}) cannot be
performed. However, both Eq. (\ref{eq:Is_expansion}) to leading order,
and the dynamical equation (\ref{eq:process_ztile}) do not depend
on $\tilde{m}$, $w$ and $\epsilon$. These two equations then fix
$\delta\hat{C}_{+}(\tilde{s})$ to lowest order near the critical
point. Given $\delta\hat{C}_{+}(\tilde{s})$, we are then left with
three equations Eqs. (\ref{eq:var_rescaled},\ref{eq:amplitude},\ref{eq:mean_rescaled})
to solve for $\tilde{m}$, $w$ and $\epsilon$ near the transition.
At the transition, $\sigma=\sqrt{2}$, $\tilde{m}=0$, $w=\sqrt{\pi}/\mu$
and $\epsilon=0$. For $\sigma=\sqrt{2}+x$ with $0<x\ll1$, we expand
close to the transition
\[
\tilde{m}=m_{1}x+m_{2}x^{2}+\dots\,,
\]
and
\[
w=\frac{\sqrt{\pi}}{\mu}\left(1+w_{1}x+\dots\right)\,,
\]
and lastly
\begin{equation}
\epsilon=\epsilon_{1}x+\dots\label{eq:exp_fluct_aging}
\end{equation}
The form of the expansion for the amplitude $\epsilon$ of the fluctuations
in Eq. (\ref{eq:exp_fluct_aging}) is imposed by Eq. (\ref{eq:amplitude}),
recalling that $\sigma=\sqrt{2}+x$. It turns out that the first corrections
to $\tilde{m}$ and $w$ can be derived explicitly from Eqs. (\ref{eq:var_rescaled},\ref{eq:mean_rescaled})
alone. We find that $\tilde{m}$ agrees with the analytical continuation
of the fixed point branch \cite{bunin2017ecological} up to order
$O\left(x^{2}\right)$ with
\begin{equation}
m_{1}=-\frac{\pi}{2\mu}\,\,\,\,\:{\rm and\,\,\,\,\:m_{2}=\frac{-4\pi^{2}+10\pi\mu-3\pi^{2}\mu}{16\sqrt{2}\mu^{2}}}\,,\label{eq:expansion_m}
\end{equation}
while $w$ agrees with the analytical continuation of the fixed point
branch up to order $O\left(x\right)$ with
\begin{equation}
w_{1}=\frac{2\pi-2\mu+\pi\mu}{2\sqrt{2}\mu}\,.\label{eq:expansion_w}
\end{equation}
To lowest order, Eq. (\ref{eq:amplitude}) gives the expression of
the amplitude of fluctuations close to the critical point
\begin{equation}
\epsilon_{1}=\frac{\pi}{2\mu\left(I(\infty)-I(0)\right)}\,.\label{eq:expansion_esp}
\end{equation}
Since $Q=\epsilon^{2}$, this equation yields the critical exponent
$\beta=2$ as defined through Eq. (\ref{eq:defQc}). The expression
for $\epsilon_{1}$ is not explicit, as it relies of the solution
of the nonlinear equation Eq. (\ref{eq:self_consdC_l0+}) for the
correlation function, so $Q_{c}(\lambda,\mu)$ in Eq. (\ref{eq:defQc})
is not given explicitly.. The rescaling of time used to obtain Eq.
(\ref{eq:self_consdC_l0+}) entails the scaling exponent $\zeta=\beta/2=1$.
As we show below, the scaling relation $\zeta=\beta/2$ is common
to both the finite $\lambda>0$ and $\lambda\to0^{+}$ critical regimes.

\noindent The picture behind these scaling results is the following:
when the temporal fluctuations in $\xi(t)$ are small, of order $\epsilon$,
only species whose time-averaged growth rate, $1-\mu m+\sigma\overline{\xi_{i}(t)}$,
is $O(\epsilon)$ show any significant fluctuations in their $\ln N_{i}$
values. The instantaneous growth rates of those that do is of order
$\epsilon$ (negative or positive), and so the exponential growth
and decline of $N_{i}$ between $O(\lambda)$ and $O(1)$ (see Fig.
\ref{fig:small_l_fig}) , takes time of order $\left|\ln\lambda\right|/\epsilon$.
As, by definition, $Q\sim\epsilon^{2}$ this explains the scaling
relation $\beta=\zeta/2$.

\subsection{Steady-state dynamics when $\lambda=0$\label{subsec:Mapping-nomig}}

\begin{figure}
\begin{centering}
\includegraphics[width=0.8\columnwidth]{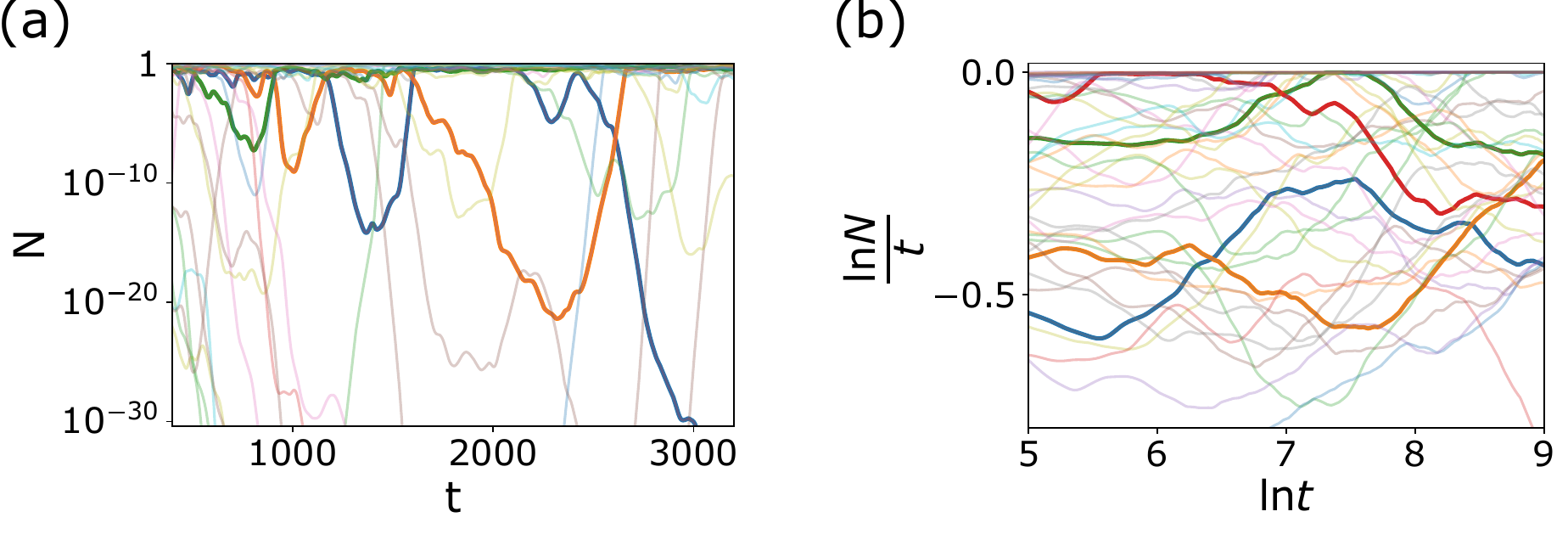}
\par\end{centering}
\caption{\label{fig:aging_fig}\textbf{Aging dynamics and time-translation
invariant dynamics of the process $z(s)$.} Example degrees of freedom
from a simulation of the dynamics Eq. (\ref{eq:many_bodyLV}) with
$\lambda=0$, $\mu=10$, $\sigma-\sigma_{c}=0.4$ and $S=4000$.\textbf{
(a) }Growth of the relaxation time and amplitude of the fluctuations
of log-population sizes with the elapsed time in the aging regime.
\textbf{(b)} Corresponding time-translation invariant process $z(s)$
where $s=\ln t$ and $z=\ln N/t$.}
\end{figure}

When $\lambda=0$, the long-time dynamics can be described in a similar
way using a Lamperti transformation of the original equations of motion
Eq. (\ref{eq:DMFT_redef}), see Fig. \ref{fig:aging_fig}. We introduce
$z=\ln N/t$ and $s=\ln t$. When $t\to\infty$, the process $z(s)$
follows a well-defined stochastic differential equation \cite{de2024many},
\begin{equation}
z'(s)=-z(s)+\sigma g+\sigma\epsilon\delta\hat{\xi}(s)+W(z)\,,\label{eq:EOM_z_0}
\end{equation}
confined to $z\leq0$ and with with $\delta\hat{\xi}(s)\equiv\delta\xi(t)$
a zero-mean Gaussian noise with correlations $\delta\hat{C}_{0}(s)$.
The expression for the population size when $\lambda=0$ remains the
same as in Sec. \ref{subsec:Mapping-l0+}, with
\[
N(s)=\sigma\,\Theta(z(s))\left(g+\epsilon\delta\hat{\xi}(s)\right)\,.
\]
Therefore, the self-consistency conditions Eqs. (\ref{eq:self-cons_longtime1},\ref{eq:self-cons_longtime2})
also hold upon replacing $\delta\hat{C}_{+}(s)$ by $\delta\hat{C}_{0}(s)$.
The $\lambda=0$ and $\lambda\to0^{+}$ long-time effective dynamics
differ by the nature of the process $z(s)$: when $\lambda=0$, the
process is confined on the negative side by a harmonic potential whereas
it is confined by another hard wall at $z=-1$ when $\lambda\to0^{+}$.
This seemingly innocuous difference entails a profound distinction
when it comes to critical slowing down. However, the formal resemblance
between the two dynamics allows us to investigate their critical property
in a similar way.

\subsection{Growth of fluctuations when $\lambda=0$\label{sec:Growth-of-fluctuations-2}}

\noindent When $\lambda=0$, the calculation proceeds in a similar
way, with yet a crucial difference. When $\lambda\to0^{+}$, we obtained
to leading order an $\epsilon$-independent equation for the correlation
function by introducing the rescaled process $\tilde{z}(\tilde{s})\equiv z(\tilde{s}/\epsilon)$.
Here one achieves a similar conclusion by rescaling $z$ following
$\tilde{z}(s)\equiv z(s)/\epsilon$. From Eq. (\ref{eq:EOM_z_0}),
we find that the latter obeys
\[
\frac{{\rm d}\tilde{z}}{{\rm d}s}=-z+\sigma\left(\tilde{g}+\delta\xi(s)\right)-W(\tilde{z})\,.
\]
The rest follows as before and to leading order, 
\begin{equation}
\delta\hat{C}_{0}(s)=\frac{I(\infty)-I(s)}{I(\infty)-I(0)}\,,\label{eq:dc_l0}
\end{equation}
with 
\[
I(s)=\int_{-\infty}^{+\infty}{\rm d}\tilde{g}\left\langle \left[\Theta(\tilde{z}(s))\Theta(\tilde{z}(0))-\Theta(\tilde{g})\right]\left(\tilde{g}+\delta\xi(0)\right)\left(\tilde{g}+\delta\xi(s)\right)\right\rangle _{\tilde{g}}\,.
\]
The expansion around $\sigma=\sqrt{2}$ of $\tilde{m}$, $w$ and
$\epsilon$ remains the same. Thus Eqs. (\ref{eq:expansion_m},\ref{eq:expansion_w},\ref{eq:expansion_esp})
remain valid and the exponent $\beta=2$ is preserved. The absence
of rescaling of time to obtain Eq. (\ref{eq:dc_l0}) entails the scaling
exponent $\zeta=0$, meaning that there is no critical slowing down
of the log-time dynamics.

\section{Finite $\lambda$ criticality\label{sec:finite_l}}

In this section, we investigate the properties of the critical point
when the migration rate $\lambda>0$ is finite. The amplitude of the
fluctuations vanishes at the critical point, so that $\epsilon\to0$.
Furthermore, we will show that the correlation-time of $\delta\xi(t)$
diverges at the critical point, see Eq. (\ref{eq:correl_lpos}), a
form of critical slowing down. Hence, near the transition $\delta\xi(t)$
changes slowly in time. We thus perform an ``adiabatic'' expansion
of $N(t)$, around the fixed point value of Eq. (\ref{eq:DMFT_redef})
that it would reach if $\delta\xi(t)$ was constant in time, which
is
\begin{equation}
\bar{N}(t)=\frac{\sigma\left(g+\epsilon\delta\xi(t)\right)+\sqrt{\sigma^{2}\left(g+\epsilon\delta\xi(t)\right)^{2}+4\lambda}}{2}\,.\label{eq:defNbar}
\end{equation}
We now introduce the deviation $\delta N(t)=N(t)-\bar{N}(t)$. We
anticipate that $\delta N(t)$ is small close to the transition and
define $\widehat{\delta N}(t)\equiv\delta N(t)/\epsilon$. Differentiating
the dynamical mean-field equation Eq. (\ref{eq:DMFT_redef}), yields
exactly

\begin{equation}
\frac{{\rm {\rm d}\widehat{\delta N}(t)}}{{\rm d}t}=-\left(\sqrt{\sigma^{2}\left(g+\epsilon\delta\xi(t)\right)^{2}+4\lambda}+\epsilon\,\widehat{\delta N}(t)\right)\widehat{\delta N}(t)-\frac{\sigma}{2}\left(1+\frac{\sigma\left(g+\epsilon\delta\xi(t)\right)}{\sqrt{\sigma^{2}\left(g+\epsilon\delta\xi(t)\right)^{2}+4\lambda}}\right)\dot{\delta\xi}(t)\,.\label{eq:def-hat-N-1}
\end{equation}
We can now rewrite the self-consistency conditions Eqs. (\ref{eq:DMFT1-1},\ref{eq:DMFT2})
in terms of the processes $\widehat{\delta N}(t)$ and $\bar{N}(t)$,
which will be at the basis of our expansion close to the critical
point. Lastly, it is known that at long times the dynamics in Eq.
(\ref{eq:DMFT_redef}) reaches a time-translation invariant state
\cite{roy2019numerical}, so that one can replace $\delta C(t,t+\tau)$
by $\delta C(\tau)$. This leads to the self-consistency equations
\begin{equation}
\frac{1-\sigma\tilde{m}}{\mu}-\int_{-\infty}^{+\infty}{\rm d}g\,P(g)\left\langle \bar{N}(t)\right\rangle _{g}=\epsilon\int_{-\infty}^{+\infty}{\rm d}g\,P(g)\left\langle \widehat{\delta N}(t)\right\rangle _{g}\,,\label{eq:DMFTmean2}
\end{equation}
and
\begin{align}
&w^{2}+\epsilon^{2}\delta C(t)-\int_{-\infty}^{+\infty}{\rm d}g\,P(g)\left\langle \bar{N}(t)\bar{N}(0)\right\rangle _{g} =\epsilon\int_{-\infty}^{+\infty}{\rm d}g\,P(g)\left\langle \bar{N}(0)\widehat{\delta N}(t)\right\rangle _{g} \nonumber \\ & +\epsilon\int_{-\infty}^{+\infty}{\rm d}g\,P(g)\left\langle \bar{N}(t)\widehat{\delta N}(0)\right\rangle _{g}+\epsilon^{2}\int_{-\infty}^{+\infty}{\rm d}g\,P(g)\left\langle \widehat{\delta N}(0)\widehat{\delta N}(t)\right\rangle _{g}\,.\label{eq:DMFTvar2}
\end{align}
We introduce the notation
\[
\mathcal{\chi}_{\tilde{m},w}\left(\epsilon^{2}\right)\equiv\int_{-\infty}^{+\infty}{\rm d}g\,P(g)\left\langle \bar{N}(t)\right\rangle _{g}\,,
\]

\noindent which depends on the parameters $\tilde{m}$, $w$ and is
regarded as a function of $\epsilon^{2}$. We also introduce

\[
\mathcal{H}_{\tilde{m},w}\left(x=\epsilon^{2},y=\epsilon^{2}\delta C(t)\right)\equiv\int_{-\infty}^{+\infty}{\rm d}g\,P(g)\left\langle \bar{N}(t)\bar{N}(0)\right\rangle _{g},
\]
which depends on the parameters $\tilde{m}$, $w$ and is regarded
as a function of the two variables $x=\epsilon^{2}$ and $y=\epsilon^{2}\delta C(t)$.
The fact that $\delta\xi(t)$, appearing in the definition of $\bar{N}(t)$
in Eq. (\ref{eq:defNbar}), is Gaussian guarantees the existence of
this functional form. Lastly, we also introduce 
\begin{align}
\mathcal{I}_{\tilde{m},w}[\delta C](t)& \equiv\epsilon^{-1}\int_{-\infty}^{+\infty}{\rm d}g\,P(g)\left\langle \bar{N}(0)\widehat{\delta N}(t)\right\rangle _{g}+\epsilon^{-1}\int_{-\infty}^{+\infty}{\rm d}g\,P(g)\left\langle \bar{N}(t)\widehat{\delta N}(0)\right\rangle _{g} \nonumber \\ & +\int_{-\infty}^{+\infty}{\rm d}g\,P(g)\left\langle \widehat{\delta N}(0)\widehat{\delta N}(t)\right\rangle _{g}\,,
\end{align}
which is a functional of the correlation function. It therefore depends
on $\delta C(s)$ for all $0\leq s\leq t$. With these notations we
can rewrite Eq. (\ref{eq:DMFTvar2}) in a compact way
\begin{equation}
w^{2}+\epsilon^{2}\delta C(t)-\mathcal{H}_{\tilde{m},w}\left(\epsilon^{2},\epsilon^{2}\delta C(t)\right)=\epsilon^{2}\mathcal{I}_{\tilde{m},w}[\delta C](t)\,.\label{eq:rewritting_eqC}
\end{equation}

\noindent This allows to obtain a compact equation satisfied by the
correlation function $\delta C(t)$ as follows. First, when $t\to\infty$,
we have by definition $\delta C(t)\to0$. Hence
\begin{equation}
w^{2}=\mathcal{H}_{\tilde{m,}w}\left(\epsilon^{2},0\right)+\epsilon^{2}\mathcal{I}_{\tilde{m,}w}[\delta C](\infty)\,.\label{eq:correl_ezeor}
\end{equation}
Therefore,
\[
\epsilon^{2}\delta C(t)-\left(\mathcal{H}_{\tilde{m},w}\left(\epsilon^{2},\epsilon^{2}\delta C(t)\right)-\mathcal{H}_{\tilde{m},w}\left(\epsilon^{2},0\right)\right)=\epsilon^{2}\left(\mathcal{I}_{\tilde{m},w}[\delta C](t)-\mathcal{I}_{\tilde{m},w}[\delta C](\infty)\right)\,,
\]
which, in the chaotic phase where $\epsilon>0$, simplifies to
\begin{equation}
\delta C(t)-\frac{\mathcal{H}_{\tilde{m},w}\left(\epsilon^{2},\epsilon^{2}\delta C(t)\right)-\mathcal{H}_{\tilde{m},w}\left(\epsilon^{2},0\right)}{\epsilon^{2}}=\mathcal{I}_{\tilde{m},w}[\delta C](t)-\mathcal{I}_{\tilde{m},w}[\delta C](\infty)\,.\label{eq:equation_dC}
\end{equation}
So far, all these self-consistency equations are exact. In the following,
we solve them in the vicinity of the critical point. We start by showing
that the solution $\delta C(t)$ of Eq. (\ref{eq:equation_dC}) exhibits
a diverging correlation time when $\epsilon\to0$. We then use this
fact to obtain the critical value $\sigma_{c}(\mu,\lambda)$ above
which the system enters the chaotic phase. This generalizes earlier
results (\cite{bunin2017ecological}) showing that $\sigma_{c}(\mu,\lambda)\to\sqrt{2}$
as $\lambda\to0^{+}$. We also obtain predictions for $\tilde{m}_{c}$
and $w_{c}$, the values of $\tilde{m}$ and $w$ respectively, at
the transition. Then we investigate the vicinity of the critical point,
get the critical exponents and derive the solution $\delta C(t)$
of Eq. (\ref{eq:equation_dC}) to leading order when $\epsilon\to0$.
Lastly, we investigate the fate of these results when $\lambda\ll1$.

\subsection{Existence of critical slowing down}

We start by assuming that there is no critical slowing down and that
the timescales over which $\delta C(t)$ evolves remain finite when
approaching the transition from above. We will eventually show that
this ansatz does not lead to any physical solution of Eq. (\ref{eq:equation_dC}).
Let $\delta C_{0}(t)\equiv\lim_{\sigma\to\sigma_{c}}\delta C(t)$
be the correlation function to zeroth order in $\epsilon$. An equation
for $\delta C_{0}(t)$ can then be obtained from Eq. (\ref{eq:equation_dC})
as follows. As $\epsilon\to0$ (or equivalently $\sigma\to\sigma_{c})$,
we get
\begin{equation}
\delta C_{0}(t)\left(1-\partial_{y}\mathcal{H}_{\tilde{m}_{c},w_{c}}\left(0,0\right)\right)=\lim_{\epsilon\to0}\,\left(\mathcal{I}_{\tilde{m},w}[\delta C](t)-\mathcal{I}_{\tilde{m},w}[\delta C](\infty)\right)\,.\label{eq:no_solwdown}
\end{equation}
We denote by $P_{c}(g)$ the probability distribution of $g$, Eq.
(\ref{eq:distrib_g}), at the transition where $\tilde{m}$ and $w$
are replaced by their value at the critical point, $\tilde{m}_{c}$
and $w_{c}$ respectively. By definition, we have
\begin{align*}
\mathcal{I}_{\tilde{m},w}[\delta C](\tau)-\mathcal{I}_{\tilde{m},w}[\delta C](\infty) & =\epsilon^{-1}\int_{-\infty}^{+\infty}{\rm d}g\,P(g)\left[\left\langle \bar{N}(0)\widehat{\delta N}(\tau)+\bar{N}(\tau)\widehat{\delta N}(0)\right\rangle _{g}-2\left\langle \bar{N}\right\rangle _{g}\left\langle \widehat{\delta N}\right\rangle _{g}\right]\\
 & +\int_{-\infty}^{+\infty}{\rm d}g\,P(g)\left[\left\langle \widehat{\delta N}(0)\widehat{\delta N}(\tau)\right\rangle _{g}-\left\langle \widehat{\delta N}\right\rangle _{g}^{2}\right]\,.
\end{align*}
We further introduce the functions 
\begin{equation}
f(x)=\frac{x+\sqrt{x^{2}+4\lambda}}{2}\,,\label{eq:deff}
\end{equation}
so that $\bar{N}(t)=f\left(\sigma(g+\epsilon\delta\xi(t))\right)$
and 
\[
h(x)=\sqrt{x^{2}+4\lambda}\,,
\]
so that
\[
\frac{{\rm {\rm d}\widehat{\delta N}(t)}}{{\rm d}t}=-\left(h\left(\sigma(g+\epsilon\delta\xi(t))\right)+\epsilon\,\widehat{\delta N}(t)\right)\widehat{\delta N}(t)-\sigma f'\left(\sigma(g+\epsilon\delta\xi(t))\right)\dot{\delta\xi}(t)\,.
\]
We can expand the value of $\bar{N}(t)$ around $f(g)$ to get to
leading order
\begin{align*}
\epsilon^{-1}\int_{-\infty}^{+\infty}{\rm d}g\,P(g)\left[\left\langle \bar{N}(0)\widehat{\delta N}(\tau)+\bar{N}(\tau)\widehat{\delta N}(0)\right\rangle _{g}-2\left\langle \bar{N}\right\rangle _{g}\left\langle \widehat{\delta N}\right\rangle _{g}\right]\\
=\sigma\int_{-\infty}^{+\infty}{\rm d}g\,P(g)f'(\sigma g)\left[\left\langle \delta\xi(0)\widehat{\delta N}(\tau)\right\rangle _{g}+\left\langle \delta\xi(\tau)\widehat{\delta N}(0)\right\rangle _{g}\right]+O\left(\epsilon\right)\,.
\end{align*}

\noindent Hence we obtain
\begin{align*}
\lim_{\epsilon\to0}\left(\mathcal{I}_{\tilde{m,}w}[\delta C](t)-\mathcal{I}_{\tilde{m,}w}[\delta C](\infty)\right) & =\sigma\int_{-\infty}^{+\infty}{\rm d}g\,P_{c}(g)f'(\sigma g)\left[\left\langle \delta\xi(0)\widehat{\delta N}(\tau)\right\rangle _{g}+\left\langle \delta\xi(\tau)\widehat{\delta N}(0)\right\rangle _{g}\right]\\
 & +\int_{-\infty}^{+\infty}{\rm d}g\,P_{c}(g)\left\langle \widehat{\delta N}(0)\widehat{\delta N}(t)\right\rangle _{g}\,.
\end{align*}
Therefore, to get the right-hand side of Eq. (\ref{eq:no_solwdown}),
we can replace $\delta\xi(t)$ by a zero-mean Gaussian process $\delta\xi_{0}(t)$
with correlations $\left\langle \delta\xi_{0}(t)\delta\xi_{0}(0)\right\rangle =\delta C_{0}(t)$
and we can also approximate $\widehat{\delta N}(t)$ by the process
$\widehat{\delta N}_{0}(t)$, which obeys

\noindent 
\begin{equation}
\frac{{\rm {\rm d}\widehat{\delta N}_{0}(t)}}{{\rm d}t}=-h\left(\sigma_{c}g\right)\widehat{\delta N}_{0}(t)-\sigma_{c}f'\left(\sigma_{c}g\right)\dot{\delta\xi}_{0}(t)\,,\label{eq:defn0}
\end{equation}
and which is solved at a steady-state by
\[
\widehat{\delta N}_{0}(t)=-\sigma_{c}f'\left(\sigma_{c}g\right)\int_{-\infty}^{t}{\rm d}u\:{\rm exp}\left(-(t-u)h\left(\sigma_{c}g\right)\right)\dot{\delta\xi}_{0}(u)\,.
\]
We can now write
\begin{align}
\left\langle \widehat{\delta N}_{0}(t)\widehat{\delta N}_{0}(0)\right\rangle _{g} & =-\sigma_{c}^{2}f'\left(\sigma_{c}g\right)^{2}\int_{-\infty}^{+\infty}{\rm d}u\:\frac{{\rm exp}\left(-|u|\,h(\sigma_{c}g)\right)}{2\,h(\sigma_{c}g)}\delta C_{0}''(t-u)\,,\label{eq:correldNdN}
\end{align}
together with
\begin{align}
\left\langle \delta\xi_{0}(0)\widehat{\delta N}_{0}(t)+\delta\xi_{0}(t)\widehat{\delta N}_{0}(0)\right\rangle _{g} & =-\sigma_{c}f'\left(\sigma_{c}g\right)\int_{-\infty}^{+\infty}{\rm d}u\:{\rm exp}\left(-|u|\,h(\sigma_{c}g)\right)\left(2\Theta(u)-1\right)\delta C_{0}'(t-u)\,.\label{eq:correldNdxi}
\end{align}
To leading order in the vicinity of the critical point, the correlation
function $\delta C_{0}(t)$ therefore satisfies the integro-differential
equation
\[
\left(1-\partial_{y}\mathcal{H}_{\tilde{m}_{c},w_{c}}\left(0,0\right)\right)\,\delta C_{0}(t)=\int_{-\infty}^{+\infty}{\rm d}u\:K_{1}(u)\,\delta C_{0}''(t-u)+\int_{-\infty}^{+\infty}{\rm d}u\:K_{2}(u)\,,\delta C_{0}'(t-u)\,.
\]
The kernels follow from Eqs. (\ref{eq:correldNdN},\ref{eq:correldNdxi})
and read
\[
K_{1}(u)=-\int_{-\infty}^{+\infty}{\rm d}g\,P_{c}(g)\left[\sigma_{c}^{2}f'\left(\sigma_{c}g\right)^{2}\frac{{\rm exp}\left(-|u|\,h(\sigma_{c}g)\right)}{2\,h(\sigma_{c}g)}\right]\,,
\]
and

\[
K_{2}(u)=-\int_{-\infty}^{+\infty}{\rm d}g\,P_{c}(g)\left[\sigma_{c}^{2}f'\left(g\right)^{2}{\rm exp}\left(-|u|\,h(\sigma_{c}g)\right)\left(2\Theta(u)-1\right)\right]\,.
\]
Noting that $K_{2}(u)=-2K_{1}'(u)$, integrating by parts yields
\begin{align}
\left(1-\partial_{y}\mathcal{H}_{\tilde{m}_{c},w_{c}}\left(0,0\right)\right)\,\delta C_{0}(t) & =\frac{1}{2}\int_{-\infty}^{+\infty}{\rm d}u\:K_{2}'(u)\,\delta C_{0}(t-u)\,.\label{eq:incons}
\end{align}
The above equation does not admit any non-trivial solution, and thus
signals a failure in our ansatz. In fact, our analysis rests on the
assumption that $\widehat{\delta N}_{0}(t)$ is of order $O(1)$,
which holds only if $\dot{\delta\xi}_{0}(t)$ is itself of $O(1)$,
see for instance Eq. (\ref{eq:defn0}) and recall that $h(g)>0$.
Hence, the inconsistency of Eq. (\ref{eq:incons}) shows that $d(\delta\xi(t))/dt$
must vanish close to the transition, or equivalently that the correlation
time must diverge at the transition. This also rules out the possibility
of a two time-step decay of the correlation function $\delta C_{0}(t)$,
with a fast $O(1)$ timescale and a slow diverging timescale. Indeed,
a similar equation would then be obtained for the fast modes only.
Note that a diverging timescale in $\delta C_{0}(t)$ implies that
the right-hand side vanishes at the transition, giving
\begin{equation}
1-\partial_{y}\mathcal{H}_{\tilde{m}_{c},w_{c}}\left(0,0\right)=0\,,\label{eq:onset}
\end{equation}
which allows us to localize the transition, as we discuss next.

\subsection{Fixed point phase and the onset of chaos}

Equation (\ref{eq:onset}) gives us the onset of chaos. In fact, for
$\sigma\leq\sigma_{c}(\mu,\lambda)$, the system reaches a fixed point
and $\epsilon=0$. Thus, for any value of $\sigma$ in this range,
one can get the mean and variance of the population sizes in the fixed
point phase using Eqs. (\ref{eq:DMFTmean2},\ref{eq:DMFTvar2}). In
terms of $\tilde{m}$ and $w$ these equations are written as
\begin{equation}
\frac{1-\sigma\tilde{m}}{\mu}-\mathcal{\chi}_{\tilde{m},w}\left(0\right)=\frac{1-\sigma\tilde{m}}{\mu}-\int_{-\infty}^{+\infty}{\rm d}g\,P(g)\left(\frac{\sigma g+\sqrt{\sigma^{2}g^{2}+4\lambda}}{2}\right)=0\,,\label{eq:FPm}
\end{equation}
and
\begin{equation}
w^{2}-\mathcal{H}_{\tilde{m},w}\left(0,0\right)=w^{2}-\int_{-\infty}^{+\infty}{\rm d}g\,P(g)\left(\frac{\sigma g+\sqrt{\sigma^{2}g^{2}+4\lambda}}{2}\right)^{2}=0\,,\label{eq:FPw}
\end{equation}
where $P(g)$ was defined in Eq. (\ref{eq:distrib_g}). At the critical
point, Eq. (\ref{eq:onset}) implies that
\begin{equation}
1-\partial_{y}\mathcal{H}_{\tilde{m}_{c},w_{c}}\left(0,0\right)=1-\sigma_{c}^{2}\int_{-\infty}^{+\infty}{\rm d}g\,P_{c}(g)\left(\frac{g\sigma_{c}+\sqrt{4\lambda+g^{2}\sigma_{c}^{2}}}{2\sqrt{4\lambda+g^{2}\sigma_{c}^{2}}}\right)^{2}=0,\label{eq:onset_chaos}
\end{equation}

\noindent and provides a third equation. It therefore allows to determine
$\tilde{m}_{c}$, $w_{c}$ and the position of the critical point
$\sigma_{c}$ when combined with Eqs. (\ref{eq:FPm},\ref{eq:FPw}).

\subsection{Critical regime}

We have proven that the correlation time diverges at the transition
and we have obtained the position of the critical point. We are now
in position to investigate the near-critical regime. We recall the
evolution
\begin{equation}
\frac{{\rm {\rm d}\widehat{\delta N}(t)}}{{\rm d}t}=-\left(h\left(\sigma(g+\epsilon\delta\xi(t))\right)+\epsilon\,\widehat{\delta N}(t)\right)\widehat{\delta N}(t)-\sigma f'\left(\sigma(g+\epsilon\delta\xi(t))\right)\dot{\delta\xi}(t)\,.\label{eq:hatN}
\end{equation}
We have seen that $\delta\xi$ must evolve on timescales that diverge
close to the transition. If the timescale diverges as $\epsilon^{-r}$,
let us define $\tau=\epsilon^{r}t$. The yet unknown exponent $r>0$
is such that $\delta\xi'(\tau)=\epsilon^{-r}\dot{\delta\xi}(t=\tau\epsilon^{-r})$
is of order $O\left(\epsilon^{0}\right)$. Later we will prove that
$r=1$, thus entailing the scaling relation $\beta=2\zeta$ relating
the critical exponents introduced in Sec. \ref{sec:Summary-of-the}.
Upon performing an additional rescaling $\widehat{\delta N}(\tau)\to\epsilon^{r}\widehat{\delta N}(\tau)$,
we get
\begin{equation}
\epsilon^{r}\widehat{\delta N}'(\tau)=-\left(h\left(\sigma(g+\epsilon\delta\xi(\tau))\right)+\epsilon^{r+1}\,\widehat{\delta N}(\tau)\right)\widehat{\delta N}(\tau)-\sigma f'\left(\sigma(g+\epsilon\delta\xi(\tau))\right)\delta\xi'(\tau)\,.\label{eq:EOM_rescaled}
\end{equation}
We can now return to the self-consistency equations, which we recall
here for clarity. First, Eqs. (\ref{eq:DMFTmean2},\ref{eq:correl_ezeor}),
which are the analogues of Eqs. (\ref{eq:FPm},\ref{eq:FPw}) in the
chaotic phase, allow to derive the static mean and variance of the
population sizes $\tilde{m}$ and $w$, at given parameters $\mu,\sigma$
and provided that $\epsilon$ and $\delta C(\tau)$ are known. They
read
\begin{equation}
\frac{1-\sigma\tilde{m}}{\mu}-\mathcal{\chi}_{\tilde{m},w}\left(\epsilon^{2}\right)=\epsilon^{r+1}\int_{-\infty}^{+\infty}{\rm d}g\,P(g)\left\langle \widehat{\delta N}(\tau)\right\rangle _{g}\,,\label{eq:eqm}
\end{equation}
and
\begin{equation}
w^{2}=\mathcal{H}_{\tilde{m,}w}\left(\epsilon^{2},0\right)+\epsilon^{2}\mathcal{I}_{\tilde{m,}w}[\delta C](\infty)\,,\label{eq:eqw}
\end{equation}
Then, Eq. (\ref{eq:equation_dC}) provides an equation for the correlation
function $\delta C(\tau)$ itself,
\begin{equation}
\delta C(\tau)-\frac{\mathcal{H}_{\tilde{m},w}\left(\epsilon^{2},\epsilon^{2}\delta C(\tau)\right)-\mathcal{H}_{\tilde{m},w}\left(\epsilon^{2},0\right)}{\epsilon^{2}}=\mathcal{I}_{\tilde{m},w}[\delta C](\tau)-\mathcal{I}_{\tilde{m},w}[\delta C](\infty)\,.\label{eq:eqdC}
\end{equation}
The requirement that there exists a solution to Eq. (\ref{eq:eqdC})
satisfying $\delta C(\tau)=1,\,\delta C(\infty)=0$ together with
$\delta C'(0)=0$, meaning that the correlation function is regular
at $\tau=0$, imposes the value of $\epsilon$. We can now evaluate
the scaling in $\epsilon$ of the different terms entering the self-consistency
equations.

\subsubsection{Scaling analysis of Eq. (\ref{eq:EOM_rescaled})}

We start by analyzing Eq. (\ref{eq:EOM_rescaled}) and compute the
moments of the process $\widehat{\delta N}(\tau)$ entering the self-consistency
equations. Because these self-consistency conditions relate the correlations
of $\delta\xi(\tau)$ and $\delta N(\tau)$, we get that $\widehat{\delta N}'(\tau)$
is also of order $O\left(\epsilon^{0}\right)$. To leading order,
the dynamics Eq. (\ref{eq:EOM_rescaled}) thus reduces to 
\begin{equation}
\widehat{\delta N}(\tau)=-\sigma\frac{f'\left(\sigma g\right)}{h\left(\sigma g\right)}\delta\xi'(\tau)+O(\epsilon)+O\left(\epsilon^{r}\right)\,,\label{eq:LOdN}
\end{equation}

\noindent which implies
\begin{equation}
\left\langle \widehat{\delta N}(0)\widehat{\delta N}(\tau)\right\rangle _{g}=\left(\sigma\frac{f'\left(\sigma g\right)}{h\left(\sigma g\right)}\right)^{2}\left\langle \delta\xi'(\tau)\delta\xi'(0)\right\rangle +O\left(\epsilon\right)+O\left(\epsilon^{r}\right)\,.\label{eq:LOvardN}
\end{equation}

\noindent The mean value of $\widehat{\delta N}(\tau)$ in steady-state
vanishes to order $O(1)$. In fact, it also vanishes to order $O(\epsilon)$.
Indeed, we remark that Eq. (\ref{eq:EOM_rescaled}) yields

\[
\left\langle \widehat{\delta N}(\tau)h\left(g+\epsilon\delta\xi(\tau)\right)\right\rangle _{g}+\epsilon^{r+1}\,\left\langle \widehat{\delta N}(\tau)\widehat{\delta N}(\tau)\right\rangle _{g}=-\left\langle \sigma f'\left(\sigma(g+\epsilon\delta\xi(\tau))\right)\delta\xi'(\tau)\right\rangle \,.
\]
Therefore,
\begin{align}
& h\left(\sigma g\right)\left\langle \widehat{\delta N}(\tau)\right\rangle _{g}+\epsilon\sigma\,h'\left(\sigma g\right)\left\langle \widehat{\delta N}(\tau)\delta\xi(\tau)\right\rangle _{g}+\epsilon^{r+1}\,\left\langle \widehat{\delta N}(\tau)\widehat{\delta N}(\tau)\right\rangle _{g} \nonumber \\ & =-\epsilon\,\sigma^{2}f''\left(\sigma g\right)\left\langle \delta\xi(\tau)\delta\xi'(\tau)\right\rangle +O\left(\epsilon^{2}\right)\,,
\end{align}
which, using Eq. (\ref{eq:LOdN}), gives
\begin{equation}
\left\langle \widehat{\delta N}(\tau)\right\rangle _{g}=O\left(\epsilon^{2}\right)+O\left(\epsilon^{r+1}\right)\,,\label{eq:LOmeandN}
\end{equation}

\noindent because $\left\langle \delta\xi(\tau)\delta\xi'(\tau)\right\rangle =\delta C'(0)=0$
since $\delta C(\tau)$ is an even function. For reasons that will
become clear below, we lastly need to evaluate $\left\langle \delta\xi(0)\widehat{\delta N}(\tau)+\delta\xi(\tau)\widehat{\delta N}(0)\right\rangle _{g}$.
Upon rewriting Eq. (\ref{eq:EOM_rescaled}), we get
\begin{equation}
\widehat{\delta N}(\tau)=-\frac{\epsilon^{r}\widehat{\delta N}'(\tau)}{h\left(\sigma(g+\epsilon\delta\xi(\tau))\right)}-\frac{\epsilon^{r+1}}{h\left(\sigma(g+\epsilon\delta\xi(\tau))\right)}\,\widehat{\delta N}(\tau)\widehat{\delta N}(\tau)-\sigma\frac{f'\left(\sigma(g+\epsilon\delta\xi(\tau))\right)}{h\left(\sigma(g+\epsilon\delta\xi(\tau))\right)}\delta\xi'(\tau)\,.\label{eq:dNLObis}
\end{equation}

\noindent We proceed term by term. First we have
\begin{align}
& \left\langle \delta\xi(0)\frac{\epsilon^{r}\widehat{\delta N}'(\tau)}{h\left(\sigma(g+\epsilon\delta\xi(\tau))\right)}\right\rangle _{g}=-\epsilon^{r}\sigma\frac{f'\left(\sigma g\right)}{h\left(\sigma g\right)^{2}}\left\langle \delta\xi(0)\delta\xi''(\tau)\right\rangle _{g}+O\left(\epsilon^{r+1}\right)+O\left(\epsilon^{2r}\right)\nonumber \\ & =-\epsilon^{r}\sigma\frac{f'\left(\sigma g\right)}{h\left(\sigma g\right)^{2}}\delta C''(\tau)+O\left(\epsilon^{r+1}\right)+O\left(\epsilon^{2r}\right)\,.
\end{align}

\noindent Then 
\[
\left\langle \delta\xi(0)\frac{\epsilon^{r+1}}{h\left(\sigma(g+\epsilon\delta\xi(\tau))\right)}\,\widehat{\delta N}(\tau)\widehat{\delta N}(\tau)\right\rangle _{g}=O\left(\epsilon^{r+2}\right)\,,
\]

\noindent as the leading order is the product of an odd number of
zero-mean Gaussian variables. Finally, the last term in Eq. (\ref{eq:dNLObis})
could appear to dominate the correlation $\left\langle \delta\xi(0)\widehat{\delta N}(\tau)+\delta\xi(\tau)\widehat{\delta N}(0)\right\rangle _{g}$
but it doesn't due to the $\tau\to-\tau$ symmetry. Indeed
\begin{align}
& \left\langle \delta\xi(0)\frac{f'\left(\sigma(g+\epsilon\delta\xi(\tau))\right)}{h\left(\sigma(g+\epsilon\delta\xi(\tau))\right)}\delta\xi'(\tau)+\delta\xi(\tau)\frac{f'\left(\sigma(g+\epsilon\delta\xi(0))\right)}{h\left(\sigma(g+\epsilon\delta\xi(0))\right)}\delta\xi'(0)\right\rangle _{g} \nonumber \\ & =\frac{f'\left(\sigma g\right)}{h\left(\sigma g\right)}\left\langle \delta\xi(0)\delta\xi'(\tau)+\delta\xi(\tau)\delta\xi'(0)\right\rangle +O\left(\epsilon^{2}\right)=O\left(\epsilon^{2}\right)\,.
\end{align}

\noindent Therefore,
\begin{equation}
\left\langle \delta\xi(0)\widehat{\delta N}(\tau)+\delta\xi(\tau)\widehat{\delta N}(0)\right\rangle _{g}=2\epsilon^{r}\sigma_{c}\frac{f'\left(\sigma_{c}g\right)}{h\left(\sigma_{c}g\right)^{2}}\delta C''(\tau)+O\left(\epsilon^{2}\right)+O\left(\epsilon^{2r}\right)\,,\label{eq:crossXidN}
\end{equation}

\noindent where to leading order we have replaced $\sigma$ by its
value at the critical point.

\subsubsection{Scaling analysis of Eq. (\ref{eq:eqdC})}

We can now analyze the small $\epsilon$ behavior of Eq. (\ref{eq:eqdC})
which requires computing $\mathcal{I}_{\tilde{m},w}[\delta C](\tau)-\mathcal{I}_{\tilde{m},w}[\delta C](\infty)$.
By definition, it reads
\begin{align*}
\mathcal{I}_{\tilde{m},w}[\delta C](\tau)-\mathcal{I}_{\tilde{m},w}[\delta C](\infty) & =\epsilon^{-1+r}\int_{-\infty}^{+\infty}{\rm d}g\,P(g)\left\langle f\left(\sigma(g+\epsilon\delta\xi(0))\right)\widehat{\delta N}(\tau)+f\left(\sigma(g+\epsilon\delta\xi(\tau))\right)\widehat{\delta N}(0)\right\rangle _{g}\\
-2\epsilon^{-1+r}\int_{-\infty}^{+\infty}{\rm d}g\,P(g) & \left\langle f\left(\sigma(g+\epsilon\delta\xi)\right)\right\rangle \left\langle \widehat{\delta N}\right\rangle _{g}+\epsilon^{2r}\int_{-\infty}^{+\infty}{\rm d}g\,P(g)\left[\left\langle \widehat{\delta N}(0)\widehat{\delta N}(\tau)\right\rangle _{g}-\left\langle \widehat{\delta N}\right\rangle _{g}^{2}\right]\,.
\end{align*}
Using Eqs. (\ref{eq:LOvardN}, \ref{eq:LOmeandN}) we first get,
\begin{align}
& \epsilon^{2r}\int_{-\infty}^{+\infty}{\rm d}g\,P(g)\left[\left\langle \widehat{\delta N}(0)\widehat{\delta N}(\tau)\right\rangle _{g}-\left\langle \widehat{\delta N}\right\rangle _{g}^{2}\right]\nonumber \\ & =\epsilon^{2r}\left\langle \delta\xi'(\tau)\delta\xi'(0)\right\rangle \int_{-\infty}^{+\infty}{\rm d}g\,P(g)\left(\sigma\frac{f'\left(\sigma g\right)}{h\left(\sigma g\right)}\right)^{2}+O\left(\epsilon^{3r}\right)+O\left(\epsilon^{1+2r}\right)\,.
\end{align}
Furthermore, we obtain 
\begin{align}
& \left\langle f\left(\sigma(g+\epsilon\delta\xi(0))\right)\widehat{\delta N}(\tau)+f\left(\sigma(g+\epsilon\delta\xi(\tau))\right)\widehat{\delta N}(0)\right\rangle -2\left\langle f\left(\sigma(g+\epsilon\delta\xi)\right)\right\rangle _{g}\left\langle \widehat{\delta N}\right\rangle _{g} \nonumber \\ & =\frac{\epsilon^{2}}{2}\sigma^{2}f''(\sigma g)\left\langle (\delta\xi(0)^{2}-1)\widehat{\delta N}(\tau)+(\delta\xi(\tau)^{2}-1)\widehat{\delta N}(0)\right\rangle _{g}\nonumber \\& +\epsilon\sigma f'(\sigma g)\left\langle \delta\xi(0)\widehat{\delta N}(\tau)+\delta\xi(\tau)\widehat{\delta N}(0)\right\rangle _{g}
+O\left(\epsilon^{3}\right)\,.
\end{align}

\noindent From Eq. (\ref{eq:crossXidN}), it appears that the first
term in the right-hand side is of order $O(\epsilon^{r+1})$. The
second term is subleading since
\[
\left\langle (\delta\xi(0)^{2}-1)\widehat{\delta N}(\tau)+(\delta\xi(\tau)^{2}-1)\widehat{\delta N}(0)\right\rangle _{g}=O\left(\epsilon\right)\,.
\]
In fact, to leading order, this expression only involves products
of an odd number of zero-mean Gaussian variables. Therefore, 
\begin{align*}
\mathcal{I}_{\tilde{m},w}[\delta C](\tau)-\mathcal{I}_{\tilde{m},w}[\delta C](\infty) & =\epsilon^{2r}\delta C''(\tau)\int_{-\infty}^{+\infty}{\rm d}g\,P(g)\left(\sigma\frac{f'\left(\sigma g\right)}{h\left(\sigma g\right)}\right)^{2}+O\left(\epsilon^{3r}\right)+O\left(\epsilon^{2+r}\right)\,,
\end{align*}

\noindent since $\left\langle \delta\xi'(\tau)\delta\xi'(0)\right\rangle =-\delta C''(\tau)$.
Finally, we expand the left-hand side of Eq. (\ref{eq:eqdC}) to obtain
\begin{align}
\delta C(\tau)-\frac{\mathcal{H}_{\tilde{m},w}\left(\epsilon^{2},\epsilon^{2}\delta C(\tau)\right)-\mathcal{H}_{\tilde{m},w}\left(\epsilon^{2},0\right)}{\epsilon^{2}} & =\delta C(\tau)\left(1-\partial_{y}\mathcal{H}_{\tilde{m},w}\left(\epsilon^{2},0\right)\right)\nonumber \\ & -\frac{1}{2}\epsilon^{2}\delta C(\tau)^{2}\partial_{y}^{2}\mathcal{H}_{\tilde{m},w}\left(0,0\right)+O\left(\epsilon^{4}\right)\,.
\end{align}

\noindent We now use the equation for the onset of chaos Eq. (\ref{eq:onset_chaos})
to get

\noindent 
\begin{align}
& \delta C(\tau)-\frac{\mathcal{H}_{\tilde{m},w}\left(\epsilon^{2},\epsilon^{2}\delta C(\tau)\right)-\mathcal{H}_{\tilde{m},w}\left(\epsilon^{2},0\right)}{\epsilon^{2}} =\delta C(\tau)\left(\partial_{y}\mathcal{H}_{\tilde{m}_{c},w_{c}}\left(0,0\right)-\partial_{y}\mathcal{H}_{\tilde{m},w}\left(\epsilon^{2},0\right)\right)\nonumber \\ & -\frac{1}{2}\epsilon^{2}\delta C(\tau)^{2}\partial_{y}^{2}\mathcal{H}_{\tilde{m}_{c},w_{c}}\left(0,0\right)+O\left(\epsilon^{4}\right)\,, \nonumber \\
& =\delta C(\tau)\left(\partial_{y}\mathcal{H}_{\tilde{m}_{c},w_{c}}\left(0,0\right)-\partial_{y}\mathcal{H}_{\tilde{m},w}\left(0,0\right)\right) -\delta C(\tau)\epsilon^{2}\partial_{x}\partial_{y}\mathcal{H}_{\tilde{m}_{c},w_{c}}\left(0,0\right) \nonumber \\ & +\frac{1}{2}\epsilon^{2}\delta C(\tau)^{2}\partial_{y}^{2}\mathcal{H}_{\tilde{m}_{c},w_{c}}\left(0,0\right)+O\left(\epsilon^{4}\right)\,.
\end{align}
Therefore we obtain the equation satisfied by the correlation function
in the vicinity of the critical point,

\begin{align*}
-\delta C(\tau)\epsilon^{2}\partial_{x}\partial_{y}\mathcal{H}_{\tilde{m}_{c},w_{c}}\left(0,0\right)+\delta C(\tau)\left(\partial_{y}\mathcal{H}_{\tilde{m}_{c},w_{c}}\left(0,0\right)-\partial_{y}\mathcal{H}_{\tilde{m},w}\left(0,0\right)\right) & -\frac{1}{2}\epsilon^{2}\delta C(\tau)^{2}\partial_{y}^{2}\mathcal{H}_{\tilde{m}_{c},w_{c}}\left(0,0\right)+O\left(\epsilon^{4}\right)\\
=\epsilon^{2r}\delta C''(\tau)\int_{-\infty}^{+\infty}{\rm d}g\,P(g)\left(\sigma\frac{f'\left(\sigma g\right)}{h\left(\sigma g\right)}\right)^{2} & +O\left(\epsilon^{3r}\right)+O\left(\epsilon^{2+r}\right)\,.
\end{align*}
In order for a physically sound solution to exist, all leading order
terms must have the same scaling as $\epsilon\to0$. This imposes
the scaling behavior $r=1$. In the following, we use the notations
\begin{align}
\gamma & =-\partial_{x}\partial_{y}\mathcal{H}_{\tilde{m}_{c},w_{c}}\left(0,0\right)+\lim_{\sigma\to\sigma_{c}}\frac{\partial_{y}\mathcal{H}_{\tilde{m}_{c},w_{c}}\left(0,0\right)-\partial_{y}\mathcal{H}_{\tilde{m},w}\left(0,0\right)}{\epsilon^{2}}\,,\nonumber \\
\kappa & =\frac{1}{2}\partial_{y}^{2}\mathcal{H}_{\tilde{m}_{c},w_{c}}\left(0,0\right)\,,\label{eq:definitions plural}\\
\omega & =\int_{-\infty}^{+\infty}{\rm d}g\,P_{c}(g)\left(\sigma_{c}\frac{f'\left(\sigma_{c}g\right)}{h\left(\sigma_{c}g\right)}\right)^{2}\,.\nonumber 
\end{align}
$\kappa$ and $\omega$ are determined by the known value of $w_{c}$
and $\tilde{m}_{c}$ at the transition. The parameter $\gamma$ remains
unknown as finding it requires to investigate the behavior of $w$
and $\tilde{m}$ close to the transition, as will be done in the next
section. To leading order, the correlation function obeys

\[
\delta C_{0}''(\tau)=\frac{\gamma}{\omega}\delta C_{0}(\tau)-\frac{\kappa}{\omega}\delta C_{0}(\tau)^{2}\,.
\]
The above equation can be seen as the classical equation of motion
of a massive particle in a cubic potential. The conditions $\delta C_{0}(0)=1$,
$\delta C_{0}'(0)=0$ and $\delta C_{0}(\infty)=0$ then constrain
the admissible value of $\gamma$ to be
\[
\gamma=\frac{2\kappa}{3},
\]
with $\kappa>0$. In terms of the known constants $\kappa$ and $\omega$,
the solution reads
\begin{equation}
\delta C_{0}(\tau)=1-{\rm Tanh^{2}\left(\frac{\sqrt{\kappa}\tau}{6\sqrt{\omega}}\right)}\,.\label{eq:result_C}
\end{equation}

\noindent The amplitude of the fluctuations $\epsilon$, and therefore
the critical exponent $\beta$, are derived from the constraint on
$\gamma$. As mentioned above, this requires to investigate the behavior
of $\tilde{m}$ and $w$ in the vicinity of the critical point, as
we do in Sec. \ref{sec:amplitude_finite_l}. Equation (\ref{eq:result_C})
completes the derivation of the scaling form in Eq. (\ref{eq:correl_lpos}).
We give the expression of the coefficient $\tau_{c}(\lambda,\mu)$
entering Eq. (\ref{eq:correl_lpos}) at the end of Sec. \ref{sec:amplitude_finite_l}.

\subsubsection{Scaling analysis of Eq. (\ref{eq:eqm}, \ref{eq:eqw})\label{sec:amplitude_finite_l}}

We recall that $\left\langle \widehat{\delta N}(\tau)\right\rangle _{g}=O\left(\epsilon^{2}\right)$,
see Eq. (\ref{eq:LOmeandN}). Neglecting all corrections beyond order
$\epsilon^{2}$, Eq. (\ref{eq:eqm}) becomes
\begin{equation}
\frac{1-\sigma\tilde{m}}{\mu}-\mathcal{\chi}_{\tilde{m},w}\left(0\right)-\epsilon^{2}\mathcal{\chi}_{\tilde{m}_{c},w_{c}}'\left(0\right)=\frac{1-\sigma\tilde{m}}{\mu}-\int_{-\infty}^{+\infty}{\rm d}g\,P(g)f(\sigma g)-\frac{1}{2}\epsilon^{2}\sigma_{c}^{2}\int_{-\infty}^{+\infty}{\rm d}g\,P_{c}(g)f''(\sigma_{c}g)=0\,.\label{eq:linearizedm}
\end{equation}

\noindent Additionally, we note that 
\[
\mathcal{I}_{\tilde{m,}w}[\delta C](\infty)=2\int_{-\infty}^{+\infty}{\rm d}g\,P(g)\left\langle \bar{N}\right\rangle _{g}\left\langle \widehat{\delta N}\right\rangle _{g}+\epsilon^{2}\int_{-\infty}^{+\infty}{\rm d}g\,P(g)\left\langle \widehat{\delta N}\right\rangle _{g}^{2}=O\left(\epsilon^{2}\right)\,.
\]

\noindent Therefore, neglecting all corrections beyond order $\epsilon^{2}$,
Eq. (\ref{eq:eqw}) becomes
\begin{equation}
w^{2}-\int_{-\infty}^{+\infty}{\rm d}g\,P(g)f(\sigma g)^{2}-\epsilon^{2}\sigma_{c}^{2}\int_{-\infty}^{+\infty}{\rm d}g\,P_{c}(g)f(\sigma_{c}g)f''(\sigma_{c}g)=0\,.\label{eq:linearizedw}
\end{equation}

\noindent Lastly, we recall the constraint $\gamma=2\kappa/3$ which
imposes
\begin{equation}
\lim_{\sigma\to\sigma_{c}}\frac{\partial_{y}\mathcal{H}_{\tilde{m}_{c},w_{c}}\left(0,0\right)-\partial_{y}\mathcal{H}_{\tilde{m},w}\left(0,0\right)}{\epsilon^{2}}=\frac{1}{3}\partial_{y}^{2}\mathcal{H}_{\tilde{m}_{c},w_{c}}\left(0,0\right)+\partial_{x}\partial_{y}\mathcal{H}_{\tilde{m}_{c},w_{c}}\left(0,0\right)\,.\label{eq:condition}
\end{equation}

\noindent We can now expand Eqs. (\ref{eq:condition},\ref{eq:linearizedm},\ref{eq:linearizedw})
in terms of the small parameter $x$ defined as the distance to the
critical point, $x\equiv\sigma-\sigma_{c}$. To linear level, $w=w_{c}+xw_{1}$
and $\tilde{m}=\tilde{m}_{c}+x\tilde{m}_{1}$, from which it is clear
that Eq. (\ref{eq:condition}) requires the scaling form $\epsilon=\epsilon_{0}\sqrt{x}$.
This entails the value of the critical exponent $\beta=1$, from which
it follows that $\zeta=1/2$. The expansion of Eqs. (\ref{eq:condition},\ref{eq:linearizedm},\ref{eq:linearizedw})
then fixes the values of $w_{1},\,\tilde{m}_{1}\,{\rm and}\,\epsilon_{0}$.
We get from Eqs. (\ref{eq:linearizedm},\ref{eq:linearizedw})
\begin{align}
-\tilde{m}_{1}\left(\frac{\sigma_{c}}{\mu}+\frac{1}{w_{c}^{2}}\int_{-\infty}^{+\infty}{\rm d}g\,P_{c}(g)(g-\tilde{m}_{c})f(\sigma_{c}g)\right) & -\frac{w_{1}}{w_{c}^{3}}\int_{-\infty}^{+\infty}{\rm d}g\,P_{c}(g)\left[(g-\tilde{m}_{c})^{2}-w_{c}^{2}\right]f(\sigma_{c}g)\nonumber \\
=\frac{\tilde{m}_{c}}{\mu}+\int_{-\infty}^{+\infty}{\rm d}g\,P_{c}(g)\,gf'(\sigma_{c}g) & +\frac{1}{2}\epsilon_{0}^{2}\sigma_{c}^{2}\int_{-\infty}^{+\infty}{\rm d}g\,P_{c}(g)f''(\sigma_{c}g)\,,\label{eq:eqmF}
\end{align}
and
\begin{align}
w_{1}\left(2w_{c}-\frac{1}{w_{c}^{3}}\int_{-\infty}^{+\infty}{\rm d}g\,P_{c}(g)\left[(g-\tilde{m}_{c})^{2}-w_{c}^{2}\right]f(\sigma_{c}g)^{2}\right)-\frac{\tilde{m}_{1}}{w_{c}^{2}}\int_{-\infty}^{+\infty}{\rm d}g\,P_{c}(g)(g-\tilde{m}_{c})f(\sigma_{c}g)^{2}\nonumber \\
=\int_{-\infty}^{+\infty}{\rm d}g\,P_{c}(g)2gf(\sigma_{c}g)f'(\sigma_{c}g)+\epsilon_{0}^{2}\sigma_{c}^{2}\int_{-\infty}^{+\infty}{\rm d}g\,P_{c}(g)f(\sigma_{c}g)f''(\sigma_{c}g)\,.\label{eq:eqwF}
\end{align}

\noindent Lastly, from Eq. (\ref{eq:condition}), we obtain,

\begin{align}
& \left(\sigma_{c}^{4}\int_{-\infty}^{+\infty}{\rm d}g\,P_{c}(g)f^{(3)}(g)f'(g)+\frac{1}{3}\sigma_{c}^{4}\int_{-\infty}^{+\infty}{\rm d}g\,P_{c}(g)f''(g)^{2}/3\right)\epsilon_{0}^{2} =-\frac{2\sigma_{c}\int_{-\infty}^{+\infty}{\rm d}g\,P_{c}(g)f'(g\sigma_{c})^{2}}{}\nonumber \\
& + \sigma_{c}^{2}\,\int_{-\infty}^{+\infty}{\rm d}g\,P_{c}(g)\left(\tilde{m}_{1}\frac{(g-\tilde{m}_{c})}{w_{c}^{2}}+w_{1}\frac{\left((g-\tilde{m}_{c})^{2}-w_{c}^{2}\right)}{w_{c}^{3}}\right)f'(g\sigma_{c})^{2} \nonumber \\ & +\sigma_{c}^{2}\int_{-\infty}^{+\infty}{\rm d}g\,P_{c}(g)2gf''(\sigma_{c}g)f'(\sigma_{c}g)\,.\label{eq:eqeF}
\end{align}

\noindent The parameter $\epsilon_{0}$ inferred from these equations
is directly related to the coefficient $Q_{c}(\lambda,\mu)$ of Eq.
(\ref{eq:defQc}) through $Q_{c}(\lambda,\mu)=\epsilon_{0}^{2}$.
It is also related to the coefficient $\tau_{c}(\lambda,\mu)$ of
Eq. (\ref{eq:correl_lpos}) which can be read from Eq. (\ref{eq:result_C})
\begin{equation}
\tau_{c}(\lambda,\mu)=\frac{6\sqrt{\omega}}{\epsilon_{0}\sqrt{\kappa}}\,,\label{eq:taucres}
\end{equation}

\noindent after recalling that $\tau=\epsilon t$.

\subsection{Suppression of the chaotic phase at large $\lambda$}

These results indicate that the chaotic phase is suppressed for large
values of $\lambda$, above a critical value $\lambda_{c}(\mu)$.
For $\lambda>\lambda_{c}(\mu)$, we find that the collective dynamics
in Eq. (\ref{eq:many_bodyLV}) directly experiences a transition from
the fixed point phase when $\sigma<\sigma_{c}(\lambda,\mu)$ to a
phase of unbounded growth when $\sigma>\sigma_{c}(\lambda,\mu)$,
see Fig. \ref{fig:supp_chaos} below. This phase of unbounded growth,
which corresponds to blow-up of population sizes, was already investigated
when the migration is low and was found to appear for large values
of $\sigma$ \cite{bunin2017ecological}. 

The critical parameter $\sigma_{c}(\lambda,\mu)$ which describes
the onset of stability loss of the fixed point solution is still given
by the solution of Eqs. (\ref{eq:FPm},\ref{eq:FPw},\ref{eq:onset_chaos}).
However, the set of linear equations prescribing the steady-state
amplitude of the fluctuations close to the critical point $\epsilon_{0}^{2}=Q_{c}(\lambda,\mu)$,
see Eqs. (\ref{eq:eqmF},\ref{eq:eqwF},\ref{eq:eqeF}), admits a
diverging solution at $\lambda=\lambda_{c}(\mu)$ and only (unphysical)
negative solutions for $\lambda>\lambda_{c}(\mu)$. 

\begin{figure}
\begin{centering}
\includegraphics[width=0.8\columnwidth]{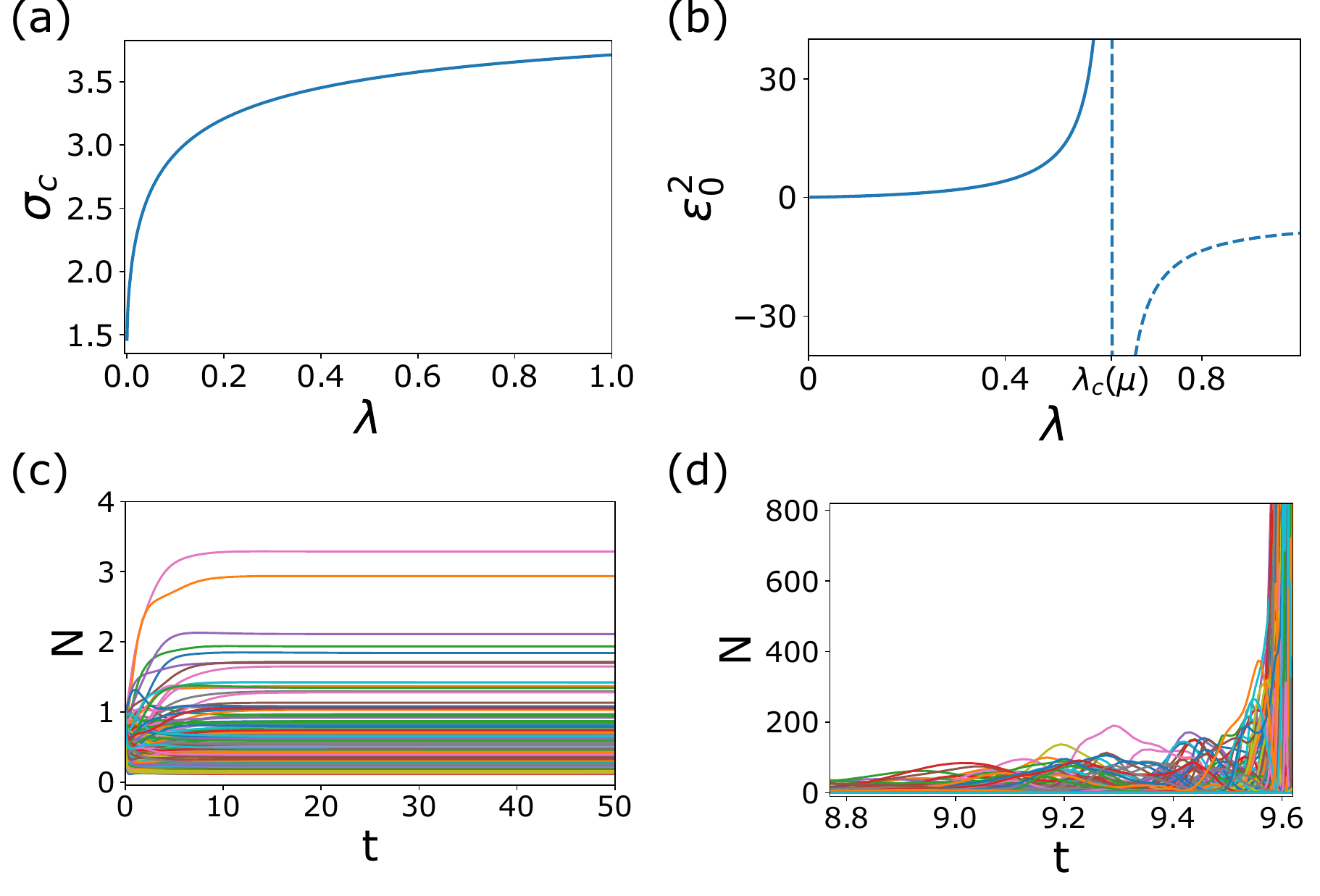}
\par\end{centering}
\caption{\label{fig:supp_chaos}\textbf{Suppression of the chaotic phase at
large migration rate.} \textbf{(a) }Onset of stability of the fixed
point phase $\sigma_{c}(\lambda,\mu)$ as a function of $\lambda$
for $\mu=10$.\textbf{ (b)} Amplitude of the chaotic fluctuations
close to the transition $Q_{c}(\lambda,\mu)=\epsilon_{0}^{2}$ as
a function of $\lambda$ for $\mu=10$ as predicted by the linear
system of Eqs. (\ref{eq:eqmF},\ref{eq:eqwF},\ref{eq:eqeF}). Above
a certain value $\lambda>\lambda_{c}(\mu)\simeq0.6$, the predicted
value for $Q_{c}(\lambda,\mu)$ is negative, therefore signaling the
suppression of the chaotic phase. \textbf{(c, d) }Many-body simulations
of Eq. (\ref{eq:many_bodyLV}) with $S=4000$ for $\mu=10$ and $\lambda=0.8>\lambda_{c}(\mu)$.
The loss of stability of the fixed phase is predicted to take place
around $\sigma_{c}(\lambda,\mu)\simeq0.65$. \textbf{(c)} Simulations
at $\sigma=0.6<\sigma_{c}(\lambda,\mu)$. The dynamics converges to
a fixed point.\textbf{ (d)} Simulations at $\sigma=0.7>\sigma_{c}(\lambda,\mu)$.
The dynamics enters a regime of unbounded growth of the population
sizes.}
\end{figure}

\subsection{Results when $\lambda\ll1$\label{sec:amplitude_small_l}}

We have characterized the near-critical regime when $\lambda>0$ is
finite. The values of $\tilde{m}_{c},\,\text{\ensuremath{w_{c}}},\,{\rm and}\,\sigma_{c}$
can be determined numerically for any finite $\lambda$ and from there
the values of $w_{1},\tilde{m}_{1}$ and $\epsilon_{0}$ can also
be determined. In the following, we focus in the regime where $\lambda\ll1$
which has been the regime of interest of most of the numerical and
analytical literature so far \cite{roy2019numerical,de2024many,dalmedigos2020dynamical,pearce2020stabilization},
and for which we make further analytical progress. Our results demonstrate
the existence of the crossover discussed in Sec. \ref{subsec:crossover},
as shown by determining $Q_{c}(\lambda,\mu)$ and $\tau_{c}(\lambda,\mu)$
when $\lambda\ll1$.

\subsubsection{Critical line}

We start with the equations yielding $\tilde{m}_{c},\,\text{\ensuremath{w_{c}}},\,{\rm and}\,\sigma_{c}$.
They read
\begin{equation}
1-\frac{\sigma_{c}^{2}}{\sqrt{2\pi w_{c}^{2}}}\int_{-\infty}^{+\infty}{\rm d}\xi\,{\rm exp}\left(-\frac{\xi^{2}}{2w_{c}^{2}}\right)f'\left(\sigma_{c}\left(\xi+\tilde{m}_{c}\right)\right){}^{2}=0\,,\label{eq:crit1l}
\end{equation}

\begin{equation}
\frac{1-\sigma\tilde{m}_{c}}{\mu}-\frac{1}{\sqrt{2\pi w_{c}^{2}}}\int_{-\infty}^{+\infty}{\rm d}\xi\,{\rm exp}\left(-\frac{\xi^{2}}{2w_{c}^{2}}\right)f\left(\sigma_{c}\left(\xi+\tilde{m}_{c}\right)\right)=0\,,\label{eq:crit2l}
\end{equation}

\begin{equation}
w_{c}^{2}-\frac{1}{\sqrt{2\pi w_{c}^{2}}}\int_{-\infty}^{+\infty}{\rm d}\xi\,{\rm exp}\left(-\frac{\xi^{2}}{2w_{c}^{2}}\right)f\left(\sigma_{c}\left(\xi+\tilde{m}_{c}\right)\right)^{2}=0\,.\label{eq:crit3l}
\end{equation}
with 
\[
f(x)=\frac{x+\sqrt{x^{2}+4\lambda}}{2}\,.
\]
We recall that $\lim_{\lambda\to0}\tilde{m}_{c}=0$ and so we rewrite
$\tilde{m}_{c}\rightarrow\sqrt{\lambda}\delta m_{c}$. We also recall
$\lim_{\lambda\to0}\sigma_{c}=\sqrt{2}$ and $\lim_{\lambda\to0}w_{c}=\sqrt{\pi}/\mu$.
This allows us to expand Eq. (\ref{eq:crit1l}) as
\begin{align*}
0 & =1-\frac{\sigma_{c}^{2}}{\sqrt{2\pi w_{c}^{2}}}\int_{-\infty}^{+\infty}{\rm d}\xi\,{\rm exp}\left(-\frac{\xi^{2}}{2w_{c}^{2}}\right)f'\left(\sigma_{c}\left(\xi+\sqrt{\lambda}\delta m_{c}\right)\right){}^{2}\,,\\
 & =1-\frac{\sigma_{c}^{2}}{2}-\frac{\sigma_{c}^{2}\sqrt{\lambda}}{\sqrt{2\pi w_{c}^{2}}}\int_{-\infty}^{+\infty}{\rm d}\xi\,{\rm exp}\left(-\frac{\lambda\,\xi^{2}}{2w_{c}^{2}}\right)\left[\frac{1}{4}\left(1+\frac{\sigma_{c}\left(\xi+\delta m_{c}\right)}{\sqrt{\sigma_{c}^{2}\left(\xi+\delta m_{c}\right)^{2}+4}}\right)^{2}-\Theta(\xi)\right]\,,\\
 & =1-\frac{\sigma_{c}^{2}}{2}-\sqrt{\lambda}\frac{\sqrt{2}\mu}{\pi}\int_{-\infty}^{+\infty}{\rm d}\xi\,\left[\frac{1}{4}\left(1+\frac{\left(\xi+\delta m_{c}\right)}{\sqrt{\left(\xi+\delta m_{c}\right)^{2}+2}}\right)^{2}-\Theta(\xi)\right]+o\left(\sqrt{\lambda}\right)\,,\\
 & =1-\frac{\sigma_{c}^{2}}{2}-\sqrt{\lambda}\frac{\sqrt{2}\mu}{\pi}\left(\delta m_{c}-\frac{\pi}{2\sqrt{2}}\right)+o\left(\sqrt{\lambda}\right)\,.
\end{align*}
We therefore expand $\sigma_{c}=\sqrt{2}+\sqrt{\lambda}\delta\sigma_{c}$
and $w_{c}=\sqrt{\pi}/\mu+\sqrt{\lambda}\delta w_{c}$ to obtain to
leading order
\begin{equation}
\delta\sigma_{c}=\frac{\mu}{\pi}\left(\frac{\pi}{2\sqrt{2}}-\delta m_{c}\right)\,.\label{eq:dcdm1}
\end{equation}
We proceed similarly to expand Eq. (\ref{eq:crit3l})
\begin{align*}
0 & =w_{c}^{2}-\frac{1}{\sqrt{2\pi w_{c}^{2}}}\int_{-\infty}^{+\infty}{\rm d}\xi\,{\rm exp}\left(-\frac{\xi^{2}}{2w_{c}^{2}}\right)f\left(\sigma_{c}\left(\xi+\tilde{m}_{c}\right)\right)^{2}\,,\\
 & =w_{c}^{2}-\frac{1}{\sqrt{2\pi w_{c}^{2}}}\int_{0}^{+\infty}{\rm d}\xi\,{\rm exp}\left(-\frac{\xi^{2}}{2w_{c}^{2}}\right)\left[\sigma_{c}^{2}\left(\xi+\sqrt{\lambda}\delta m_{c}\right)^{2}+2\lambda\right]\\
 & -\frac{\lambda^{3/2}}{\sqrt{2\pi w_{c}^{2}}}\int_{-\infty}^{+\infty}{\rm d}\xi\,{\rm exp}\left(-\frac{\lambda\xi^{2}}{2w_{c}^{2}}\right)\left[\frac{1}{4}\left(\sigma_{c}\left(\xi+\delta m_{c}\right)+\sqrt{\sigma_{c}^{2}\left(\xi+\delta m_{c}\right)^{2}+4}\right)^{2}-\left[\sigma_{c}^{2}\left(\xi+\delta m_{c}\right)^{2}+2\right]\Theta(\xi)\right]\,,\\
 & =w_{c}^{2}-\frac{1}{\sqrt{2\pi w_{c}^{2}}}\int_{0}^{+\infty}{\rm d}\xi\,{\rm exp}\left(-\frac{\xi^{2}}{2w_{c}^{2}}\right)\left[\sigma_{c}^{2}\left(\xi+\sqrt{\lambda}\delta m_{c}\right)^{2}+2\lambda\right]+O\left(\lambda^{3/2}\right)\,,\\
 & =w_{c}^{2}\left(1-\frac{\sigma_{c}^{2}}{2}\right)-\left(\sqrt{\lambda}\sqrt{\frac{2}{\pi}}\delta m_{c}w_{c}\sigma_{c}^{2}+\lambda+\frac{\lambda}{2}\sigma_{c}^{2}\delta m_{c}^{2}\right)+O\left(\lambda^{3/2}\right)\,.
\end{align*}
Therefore, to leading order, we obtain
\begin{equation}
\delta m_{c}=-\frac{\pi}{2\pi}\delta\sigma_{c}\,.\label{eq:dcdm2}
\end{equation}
Hence, Eqs. (\ref{eq:dcdm1},\ref{eq:dcdm2}) allow to find the shift
in the location of the critical point as
\begin{align*}
\delta\sigma_{c} & \simeq\frac{\mu}{\sqrt{2}}\,,\\
\delta m_{c} & \simeq-\frac{\pi}{2\sqrt{2}}\,.
\end{align*}
Lastly, $\delta w_{c}$ can be obtained to leading order by expanding
Eq. (\ref{eq:crit2l}) 
\begin{align*}
0 & =\frac{1-\sqrt{\lambda}\sigma_{c}\delta m_{c}}{\mu}-\frac{1}{\sqrt{2\pi w_{c}^{2}}}\int_{-\infty}^{+\infty}{\rm d}\xi\,{\rm exp}\left(-\frac{\xi^{2}}{2w_{c}^{2}}\right)f\left(\sigma_{c}\left(\xi+\sqrt{\lambda}\delta m_{c}\right)\right)\,\\
 & =\frac{1-\sqrt{\lambda}\sigma_{c}\delta m_{c}}{\mu}-\frac{\sigma_{c}}{\sqrt{2\pi w_{c}^{2}}}\int_{0}^{+\infty}{\rm d}\xi\,{\rm exp}\left(-\frac{\xi^{2}}{2w_{c}^{2}}\right)\left(\xi+\sqrt{\lambda}\delta m_{c}\right)\\
 & -\frac{\lambda}{\sqrt{2\pi w_{c}^{2}}}\int_{-\infty}^{+\infty}{\rm d}\xi\,{\rm exp}\left(-\frac{\lambda\xi^{2}}{2w_{c}^{2}}\right)\left[\frac{\sqrt{4+\sigma_{c}^{2}\left(\xi+\delta m_{c}\right)^{2}}}{2}-\sigma_{c}\xi\,\Theta(\xi)\right]\,.
\end{align*}
We note that the last term scales as $O\left(\lambda\ln\lambda\right)$
since
\[
\frac{\sqrt{4+\sigma_{c}^{2}\left(\xi+\delta m_{c}\right)^{2}}}{2}-\sigma_{c}\xi\,\Theta(\xi)=_{x\to\pm\infty}O\left(\frac{1}{x}\right)\,.
\]
Therefore, we obtain to leading order
\[
0=\frac{1-\sqrt{\lambda}\sigma_{c}\delta m_{c}}{\mu}-\sigma_{c}\left(\frac{w_{c}}{\sqrt{2\pi}}+\frac{\sqrt{\lambda}\delta m_{c}}{2}\right)+o\left(\sqrt{\lambda}\right)\,,
\]
which yields 
\begin{align*}
\delta w_{c} & \simeq\frac{\sqrt{\pi}}{4\mu}\left(\pi(\mu+2)-2\mu\right)\,.
\end{align*}

\subsubsection{Near-critical regime}

We now turn to the determination of $Q_{c}(\lambda,\mu)$ and $\tau_{c}(\lambda,\mu)$.
We recall that the leading order corrections close to the critical
point $\tilde{m}_{1}$, $w_{1}$ and $\epsilon_{0}$ satisfy the set
of linear equations given in Eqs. (\ref{eq:eqmF},\ref{eq:eqwF},\ref{eq:eqeF}).
After some lengthy but straightforward algebra, we obtain the leading
order expression of these coefficients. Crucially, the amplitude of
the fluctuations vanishes when $\lambda\to0$ as we find 
\begin{equation}
Q_{c}(\lambda,\mu)=\epsilon_{0}^{2}\simeq\frac{16\sqrt{\lambda}}{\sqrt{2}\mu}\,.\label{eq:qc_smalll}
\end{equation}

\noindent This suggests that the near-critical regime is controlled
by another scaling limit if the migration rate $\lambda$ goes to
$0$ faster than $\sigma$ goes to the critical value $\sigma_{c}(\lambda=0,\mu)=\sqrt{2}$,
as discussed in Sec. \ref{subsec:crossover}. It is also instructive
to look at the behavior of timescales when $\lambda\ll1$. The expression
of the timescale $\tau_{c}(\lambda,\mu)$ entering Eq. (\ref{eq:result_C})
was given in Eq. (\ref{eq:taucres}). We find that to leading order
as $\lambda\to0^{+}$, $\omega/\kappa\simeq4/3$, so that we get
\begin{equation}
\tau_{c}(\lambda,\mu)\simeq\frac{\sqrt{3\mu\sqrt{2}}}{\lambda^{1/4}}\,,\label{eq:correl_time_small}
\end{equation}

\noindent which diverges as $\lambda\to0$.

\section{Conclusion}

We have analytically described the Lotka-Volterra dynamics with many
species and random interactions between them in the vicinity of the
critical point separating the fixed phase to a phase of perpetual
fluctuations. When approaching the critical point from the fluctuating
phase, timescales are large and diverge at the transition (critical
slowing down), while the size of the temporal fluctuations decreases
continuously to zero. To characterize these two effects, we obtain
the scaling behavior of the correlation function near the critical
point. We identify two critical exponents $\beta$ and $\zeta$ in
the scaling theory, and calculate their values.

Our study highlights the effect of the migration rate $\lambda$ on
the critical dynamics. Depending on $\lambda$ and the distance from
the critical point $\sigma-\sigma_{c}$, we identify three different
scaling regimes: one for $\lambda=0$, one for $\lambda>0$ fixed,
and one for $\lambda\to0^{+}$, meaning that $\sqrt{\lambda}\ll\sigma-\sigma_{c}\ll1.$
This third regime $\sqrt{\lambda}\ll\sigma-\sigma_{c}$ is commonly
probed in numerical investigations of the dynamics \cite{roy2019numerical,dalmedigos2020dynamical}.
The scaling behavior and the values of the exponents are different
between these regimes.

This work raises a number of interesting questions for future study.
One is the study of critical behavior when approaching the transition
from the fixed point phase. In this phase there are no endogenous
dynamics at long times, but one can consider the relaxation close
to the fixed point, and the response to external noise. Previous works
have considered the linearized dynamics around the fixed point \cite{opper1992phase,roy2019numerical}
by only looking at the surviving species (those with $N>0$ at the
fixed point reached when $\lambda\to0^{+}$). Yet the present and
previous works \cite{de2024many,de2023aging} highlight the importance
of ``species turnover'' events where species are exchanged between
$O(\lambda)$ and $O(1)$ values, and this non-linear effect might
be relevant also on the fixed point side of the transition. Another
interesting direction is the behavior at finite number of species
$S$. The question of the width of the crossover region between the
fixed-point phase and the chaotic or aging ones in finite size systems
remains open, as is the behavior in this region; Simulations show
that close to the transition limit cycles are sometimes reached, even
with hundreds of variables.

Finally, it would be interesting to see if any of the critical behavior
could be observed in experiments \cite{hu2022emergent} or field
studies. The main qualitative phenomena\textendash large timescales
and small temporal fluctuations near the transition\textendash are
promising candidates.

\section*{Acknowledgements}
We thank Anna Frishman and Yanay Tovi for useful discussions.

\begin{appendix}

\section{Numerical methods\label{sec:Numerical-methods}}

Here we detail the numerical procedures used to solve the DMFT equations.
We start by giving detail about the $\lambda>0$ case. We recall the
DMFT equations 
\begin{equation}
\dot{N}(t)=N(t)\left[1-N(t)-\mu m(t)+\sigma\xi(t)\right]+\lambda\,,\label{eq:DMFT-1}
\end{equation}

\noindent with the conditions

\begin{equation}
m(t)=\left\langle N(t)\right\rangle \,,\label{eq:DMFT1-1-1}
\end{equation}

\noindent and 
\begin{equation}
C(t,t')=\left\langle \xi(t)\xi(t')\right\rangle =\left\langle N(t)N(t')\right\rangle \,,\label{eq:DMFT2-1}
\end{equation}

\noindent These are self-consistent equations. The trajectory $N(t)$
depends on $\xi(t)$, which is sampled from the correlation function
$C(t,t')$, and the function $m(t)$. Self-consistently, $C(t,t)$, $m(t)$
depend on the statistics of $N(t)$, see Eqs. (\ref{eq:DMFT1-1-1},
\ref{eq:DMFT2-1}). This self-consistency is standard in DMFT formulations.
We used a well-known numerical method to solve it \cite{eissfeller1992new,roy2019numerical}.
It starts with a guess for $C(t,t'),\,m(t),$ generates realizations
of $\xi(t)$, and from that trajectories $N(t)$, which are then used
to update $C(t,t'),\,m(t)$. This is repeated until convergence. 

In practice, at small ${\rm \delta}\sigma=\sigma-\sigma_{c}$, the
DMFT simulations were implemented with some theoretical knowledge
the expected outcome. Let $\tau_{{\rm exp}}$ be an expectation for
the correlation time in steady-state at finite ${\rm \delta}\sigma$.
Here we used $\tau_{{\rm exp}}=\tau_{c}(\lambda,\mu)/\sqrt{\delta\sigma}$.
For each iteration, the noise $\xi(t)$ was sampled from the correlation
function $C(t,t')$ over a time interval $t\in[0,100\tau_{{\rm exp}}]$.
The time interval was discretized in such a way that the bining $dt$
becomes smaller and smaller with time. Here we used $dt=0.048\tau_{{\rm exp}}$
for $t\in[0,60\tau_{{\rm exp}}]$, $dt=0.0075\tau_{{\rm exp}}$ for
$t\in[60\tau_{{\rm exp}},90\tau_{{\rm exp}}]$ and $dt=0.0033\tau_{{\rm exp}}$
for $t\in[90\tau_{{\rm exp}},100\tau_{{\rm exp}}]$. For each realization
of the dynamics in Eq. (\ref{eq:DMFT1-1}), the population size was
initialized at a near fixed point value, so that $\left(N(0)-0.01\right)\left[1-\left(N(0)-0.01\right)-\mu m(0)+\sigma\xi(0)\right]+\lambda=0$.
For the first iteration, leveraging on our estimates for fluctuations
and timescales, the initial guess for the correlation function $C(t,t')$
and mean $m(t)$ were the following: $C(t,t')=w_{c}^{2}+\delta\sigma\exp(-|t-t'|/\tau_{{\rm exp}})$
and $m(t)=m_{c}$, which are small perturbations around their values
at the critical point. We then used (i) 150 iterations with averaging
over $1\,000$ realizations and injection fraction $0.3$ followed
by (ii) 40 iterations with averaging over $10\,000$ realizations
and injection fraction $0.3$ followed by (iii) 400 iterations with
averaging over $10\,000$ realizations and injection fraction $0.03$
followed by (iv) 300 iterations with averaging over $10\,000$ realizations
and injection fraction $0.003$ followed by (v) 100 iterations with
averaging over $100\,000$ realizations and injection fraction $0.003$
and followed by (vi) 10 iterations with averaging over $1\,000\,000$
realizations and injection fraction $0.003$.

The simulations in rescaled time for the cases $\lambda\to0^{+}$
and $\lambda=0$, corresponding to the self-consistency Eqs. (\ref{eq:EOM_z_0+},\ref{eq:EOM_N_l0+},\ref{eq:self-cons_longtime1},\ref{eq:self-cons_longtime2})
and Eqs. (\ref{eq:EOM_N_l0+},\ref{eq:self-cons_longtime1},\ref{eq:self-cons_longtime2},\ref{eq:EOM_z_0})
respectively, used a very similar protocol (upon replacing $t$ by
$s$). For $\lambda>0$, we chose $\tau_{{\rm exp}}=2$ and $\tau_{{\rm exp}}=2/\delta\sigma$
for $\lambda\to0^{+}$. For the first iteration, we also chose $C(s,s')=w_{c}^{2}+\delta\sigma^{2}\exp(-|s-s'|/\tau_{{\rm exp}})$.
Lastly, the initial condition for the variable $z(s)$ at the beginning
of each realization of the dynamics was $z(s)=-1$ if $1-\mu m(0)+\sigma\xi(0)<0$
and $z(s)=0$ otherwise.

\end{appendix}

\bibliography{biblio}

\end{document}